\documentclass{article}

\PassOptionsToPackage{numbers, compress}{natbib}

\usepackage[preprint]{neurips_data_2023}

\usepackage{amsfonts}
\usepackage{amsmath}
\usepackage{booktabs}
\usepackage{float}
\usepackage[T1]{fontenc}
\usepackage{graphicx}
\usepackage{hyperref}
\usepackage[utf8]{inputenc}
\usepackage{microtype}
\usepackage{multirow}
\usepackage{nicefrac}
\usepackage{url}
\usepackage{wrapfig}
\usepackage{xcolor}
\usepackage{enumitem}

\begin{document}
    \title{VISAT: Benchmarking Adversarial and Distribution Shift Robustness in Traffic Sign Recognition with Visual Attributes}

\author{
    Simon Yu$^1$ \quad Peilin Yu$^2$ \quad Hongbo Zheng$^1$ \quad Huajie Shao$^3$ \quad Han Zhao$^1$ \quad Lui Sha$^1$ \\
    $^1$University of Illinois Urbana-Champaign \quad $^2$Brown University \quad $^3$William \& Mary \\
    \texttt{\{jundayu2,hongboz2,hanzhao,lrs\}@illinois.edu} \\
    \texttt{peilin\_yu@brown.edu} \\
    \texttt{hshao@wm.edu}
}

\maketitle

    \begin{abstract}
    We present VISAT, a novel open dataset and benchmarking suite for evaluating model robustness in the task of traffic sign recognition with the presence of \underline{vis}ual \underline{at}tributes.
    Built upon the Mapillary Traffic Sign Dataset (MTSD), our dataset introduces two benchmarks that respectively emphasize robustness against adversarial attacks and distribution shifts.
    For our adversarial attack benchmark, we employ the state-of-the-art Projected Gradient Descent (PGD) method to generate adversarial inputs and evaluate their impact on popular models.
    Additionally, we investigate the effect of adversarial attacks on attribute-specific multi-task learning (MTL) networks, revealing spurious correlations among MTL tasks.
    The MTL networks leverage visual attributes (color, shape, symbol, and text) that we have created for each traffic sign in our dataset.
    For our distribution shift benchmark, we utilize ImageNet-C's realistic data corruption and natural variation techniques to perform evaluations on the robustness of both base and MTL models.
    Moreover, we further explore spurious correlations among MTL tasks through synthetic alterations of traffic sign colors using color quantization techniques.
    Our experiments focus on two major backbones, ResNet-152 and ViT-B/32, and compare the performance between base and MTL models.
    The VISAT dataset and benchmarking framework contribute to the understanding of model robustness for traffic sign recognition, shedding light on the challenges posed by adversarial attacks and distribution shifts.
    We believe this work will facilitate advancements in developing more robust models for real-world applications in autonomous driving and cyber-physical systems.~\footnote{VISAT website: \url{http://rtsl-edge.cs.illinois.edu/visat/}}
\end{abstract}

    \section{Introduction}
\label{sec:introduction}

Traffic sign recognition plays a crucial role in ensuring the safe and reliable operation of autonomous driving systems.
Accurate detection and classification of traffic signs are vital for enabling vehicles to navigate and make informed decisions on the road.
Nevertheless, the robustness of vision models in traffic sign recognition remains a critical concern due to the presence of both model-agnostic and model-specific attacks in real-world driving scenarios.
To address this concern, we present a novel dataset and benchmarking framework that evaluates the robustness of visual models against adversarial attacks and distribution shifts in the context of traffic sign recognition with visual attributes.

Deep neural networks (DNNs), while achieving remarkable performance on various vision tasks, are susceptible to a variety of attacks.
Even slight perturbations to input images can lead to misclassifications and performance drops, posing significant risks in safety-critical applications such as autonomous driving~\cite{madry2017towards}.
Additionally, real-world driving conditions introduce model-agnostic distribution shifts caused by factors such as occlusions, lighting variations, and weather conditions~\cite{hendrycks2019benchmarking}.
These distribution shifts challenge the performance of vision models and necessitate the evaluation of their robustness in handling such shifts.

For traffic sign recognition, many prior works have primarily focused on individual aspects of robustness~\cite{hingun2022reap, zhang2022cctsdb, zhu2016traffic}, overlooking the need for a unified benchmarking framework that encompasses both model-agnostic and model-specific attacks.
In this paper, we present a comprehensive benchmarking framework that further addresses this gap by incorporating two distinct benchmarks, each targeting a specific aspect of robustness.
The first benchmark evaluates model performance under adversarial attacks, testing the resilience of vision models against carefully crafted perturbations.
The second benchmark assesses model performance under distribution shifts, simulating real-world variations and challenges in target distributions.

To facilitate our benchmarking and investigations on model robustness, we curate VISAT, an extensive open dataset that builds upon the Mapillary Traffic Sign Dataset (MTSD)~\cite{ertler2020mapillary} and incorporate additional \underline{vis}ual \underline{at}tribute labels (color, shape, symbol, and text) for each traffic sign.
While creating the visual attributes, we developed a software framework that allows for rapid labeling of the attributes, dramatically increasing the efficiency of their creation process.
The created visual attributes extend intermediate explainability and enable a more extensive evaluation with increased variety.
Using our dataset and benchmarks, we evaluate popular DNN backbones, including ResNet-152~\cite{he2016deep, he2016identity}, ViT-B/32~\cite{dosovitskiy2020image}, and, by utilizing our visual attribute labels, their MTL variants.

In addition to evaluating the robustness of models against attacks, we also investigate the impact of spurious correlations among MTL tasks in the context of traffic sign recognition using our visual attributes.
MTL models leverage a common feature extraction module to improve generalization across multiple tasks~\cite{caruana1997multitask}.
However, these tasks may exhibit spurious correlations, where models unintentionally rely on unintended and unrelated cues instead of the desired attributes~\cite{hu2022improving}.
Understanding and quantifying these spurious correlations are as crucial as assessing model robustness.

The contributions of this paper can be summarized as follows:
\begin{itemize}[leftmargin=*] 
    \setlength\itemsep{0.1em}
    \item We introduce VISAT, a novel open dataset and benchmarking suite for evaluating visual model robustness in traffic sign recognition. VISAT includes benchmarks for model-specific adversarial attacks and model-agnostic distribution shifts, providing a comprehensive evaluation.
    \item We create additional visual attribute labels (color, shape, symbol, and text) for each traffic sign, expanding VISAT's capabilities for evaluating MTL model robustness and identifying spurious correlations among MTL tasks.
    \item We conduct baseline evaluations using popular vision backbones, ResNet-152 and ViT-B/32, on our dataset, serving as benchmarks for comparing robustness against adversarial attacks and distribution shifts in traffic sign recognition.
    \item We explore the robustness of MTL models trained with our visual attributes. By cross-examining MTL task performances under various attacks, we uncover potential spurious correlations that can introduce vulnerabilities and biases affecting MTL model performance.
\end{itemize}

In the following sections, we first discuss existing works in relationship with the scopes and rationales of our work.
Subsequently, we provide detailed descriptions of the curation process of our dataset, our robustness benchmarking framework, and our analysis of the evaluation results.
Through our research, we aim to advance our understanding of model robustness in traffic sign recognition and contribute to the development of more robust and reliable models for safety-critical applications, such as autonomous driving and cyber-physical systems.

    \section{Related Work, Scopes, and Rationales}
\label{sec:related_work}

\paragraph{Traffic Sign Recognition Datasets} Traffic sign recognition has been extensively studied in the literature, leading to the creation of several datasets for training and evaluating models for recognizing traffic signs.
The German Traffic Sign Recognition Benchmark (GTSRB)~\cite{stallkamp2011german} was among the first datasets curated specifically for traffic sign classification.
Subsequently, other datasets have emerged, focusing on regional traffic signs, such as the American LISA Traffic Sign Dataset~\cite{mogelmose2012vision}, Swedish Traffic Sign Dataset~\cite{larsson2011using, larsson2011correlating}, Belgium Traffic Sign Dataset (BelgiumTS)~\cite{mathias2013traffic}, Russian Traffic Sign Dataset (RTSD)~\cite{shakhuro2016russian}, and the Chinese Tsinghua-Tencent Dataset (TT100K)~\cite{zhu2016traffic}.
While many existing traffic sign benchmarks~\cite{berghoff2021robustness, hashemi2022improving} are built upon the aged GTSRB, we employ in this study the Mapillary Traffic Sign Dataset (MTSD)~\cite{ertler2020mapillary}, standing out as a much more up-to-date and comprehensive dataset.
Unlike regional datasets, MTSD encompasses data from across the globe, offering a wider range of traffic sign variations.
Additionally, MTSD contains the largest and most diverse collection of signs, with different variants of the same sign around the world labeled as distinct classes.

\paragraph{Visual Attributes} There exists work that encodes and refines traffic sign visual attributes directly into their models and pipelines~\cite{qian2016traffic, xie2016traffic, aghdam2015unified}.
Upon our review, however, we are unable to find any existing open datasets that provide visual attributes for each sign.
Therefore, VISAT, to the best of our knowledge, is the first to directly incorporate visual attributes, such as color, shape, symbol, and text, for each sign in the dataset, providing additional explainability and meaningful information for training and evaluating MTL models, and for our robustness benchmarks and evaluations.

\paragraph{Adversarial Robustness Benchmarks} Several notable adversarial attack benchmark studies have contributed to evaluating robustness in various domains.
The RobustBench~\cite{croce2020robustbench} framework proposes a standardized benchmark for image classification, utilizing AutoAttack~\cite{croce2020reliable} to assess adversarial robustness while emphasizing the importance of evaluating various performance metrics, including distribution shifts, calibration, fairness, privacy leakage, and transferability.
REAP~\cite{hingun2022reap} introduces a large-scale benchmark specifically for adversarial patch attacks on real images, highlighting the significance of realistic geometric and lighting transformations. Additionally, there are benchmarks for low-level adversarial attacks in natural language processing (NLP)~\cite{eger2020hero}, graph machine learning (GML) models~\cite{zheng2021graph}, and facial recognition systems~\cite{goel2018smartbox}, which are not directly relevant to image-based traffic sign recognition robustness.
However, these benchmarks collectively contribute to the broader understanding of adversarial robustness across different domains.
In our work, we employ the projected gradient descent (PGD)~\cite{madry2017towards} algorithm to generate adversarial examples, which has been widely used in adversarial attack research.
By incorporating PGD-based attacks, we aim to assess the robustness of ResNet-152 and ViT-B/32 against aggressive adversarial perturbations, providing valuable insights into traffic sign adversarial robustness.

\paragraph{Distribution Shift Robustness Benchmarks} Studies such as BREEDS~\cite{santurkar2020breeds} outline the design choices for distribution shift benchmarks, including data corruption, differences in data sources, and subpopulation shifts.
In the context of data corruption, researchers have developed datasets and benchmarks to evaluate image classifiers' robustness against common corruptions and variations.
The ImageNet-C dataset and benchmarks~\cite{hendrycks2019benchmarking} offer a comprehensive evaluation of distribution shift robustness under 19 different types of data corruption techniques.
Additionally, studies such as~\cite{mintun2021interaction, michaelis2019benchmarking} explore the relationship between data augmentations, test-time corruptions, and model performance.
Furthermore, existing benchmarks of traffic sign recognition models under distribution shifts~\cite{zhang2022cctsdb, khosravian2022enhancing, zhu2016traffic, berghoff2021robustness} consider factors such as category meanings, sign sizes, weather conditions, synthetically modeled noises, illumination variations, occlusions, and geometric transformations.
In our work, we focus specifically on data corruption and variations, leveraging the ImageNet-C techniques to create our dataset and benchmarks on top of MTSD and assessing model robustness in traffic sign recognition.
Additionally, we employ our visual attributes and examine the impact of ImageNet-C corruptions on the performance of the MTL variants of ResNet-152 and ViT-B/32, allowing us to gain insights into the robustness of MTL models under realistic distribution shifts.

Extended related work can be found in Appendix~\ref{app:related_work}.

    \section{The VISAT Open Dataset}
\label{sec:dataset}

\begin{wraptable}{r}{0.4\textwidth}
    \vspace{-15pt}
    \caption{VISAT dataset composition.}
    \begin{center}
        \begin{tabular}{lll}
            \toprule
            Split       & Proportion    & Instances \\
            \midrule
            Training    & 80\%          & 242263    \\
            Validation  & 10\%          & 30283     \\
            Testing     & 10\%          & 30283     \\
            \bottomrule
        \end{tabular}
    \end{center}
    \label{tab:dataset_composition}
    \vspace{-15pt}
\end{wraptable}

\paragraph{Preprocessing} We obtained the MTSD dataset from the official Mapillary website~\cite{mapillary2022mapillary}.
MTSD annotations include object labels, bounding box sizes and locations, and other relevant details.
As our baseline experiments focus only on classification robustness, we sliced the full-frame images into sign patches using the bounding box information from the annotations, increasing overall efficiency in later processing steps and experiments.
Each of the newly generated patch images is associated with its corresponding object keys so we can later identify them. We share the release details of our codebase developed for creating the VISAT open dataset, together with hosting and licensing information, in Appendix~\ref{app:dataset}.
Unfortunately, to ensure the fairness of their own benchmarks, Mapillary intentionally left out all their official testing split annotations~\cite{ertler2020mapillary}.
To overcome this limitation, we create our own training, validation, and testing splits by randomly and uniformly sampling from the official training and validation splits.
During the splitting process, we ensured that each of the 401 classes in the dataset had at least one instance in all splits by repeating the sampling process with different seeds until this criterion was met.
The resulting dataset composition is listed under Table~\ref{tab:dataset_composition}. 

\begin{wrapfigure}{r}{0.2\textwidth}
    \vspace{-18pt}
    \begin{center}
        \includegraphics[width=0.2\textwidth]{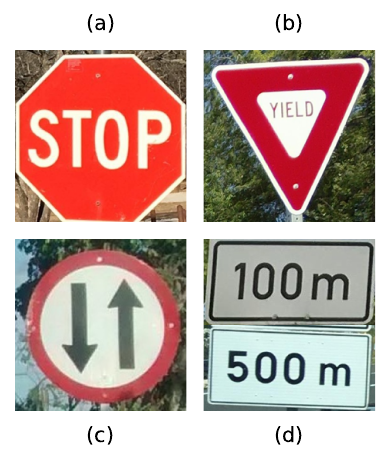}
        \vspace*{-20pt}
        \caption{Visual attribute labeling examples.}
        \label{fig:labeling_examples}
    \end{center}
    \vspace{-18pt}
\end{wrapfigure}

\paragraph{Visual Attribute Formulations} With the large total number of individual sign patches in our dataset, manually assigning visual attributes to each patch becomes infeasible and lacks generalizability.
Instead, we formulate visual attributes at the level of original class labels from the MTSD dataset.
For the 401 original classes, we observe the sign patch instances within each class and assign visual attributes based on their shared general features.
Our VISAT dataset defines 4 types of visual attributes: color, shape, symbol, and text, representing distinct and fundamental characteristics of typical traffic signs.
Color attributes capture major background and foreground colors. For example, the stop sign shown in Figure~\ref{fig:labeling_examples}-(a) is labeled as "color--red--white".
Shape attributes directly encode the shape of the signs, with additional characteristics appended if necessary, e.g., Figure~\ref{fig:labeling_examples}-(b) as "shape--triangle--inverted". Symbol attributes describe symbols present in the signs using words, with additional characteristics encoded as needed.
In some cases, we inherit categorizations from MTSD for symbols with subtle differences. Multiple symbols in a sign, e.g., the reciprocal arrows in Figure~\ref{fig:labeling_examples}-(c), are described in their geometric order.
Text attributes directly embed the texts written on the signs.
On certain occasions, signs of the same type with different texts are grouped together by MTSD under a single class label.
For instance, sets of distance signs, as shown in Figure~\ref{fig:labeling_examples}-(d), despite describing a variety of distances, are grouped in 1 class.
In this case, we label the text attributes of these signs as "text--alphanumeric". The extended statistics and information on the VISAT visual attributes can be found in Appendix~\ref{app:dataset}.

\begin{wrapfigure}{r}{0.4\textwidth}
    \vspace{-15pt}
    \begin{center}
        \includegraphics[width=0.38\textwidth]{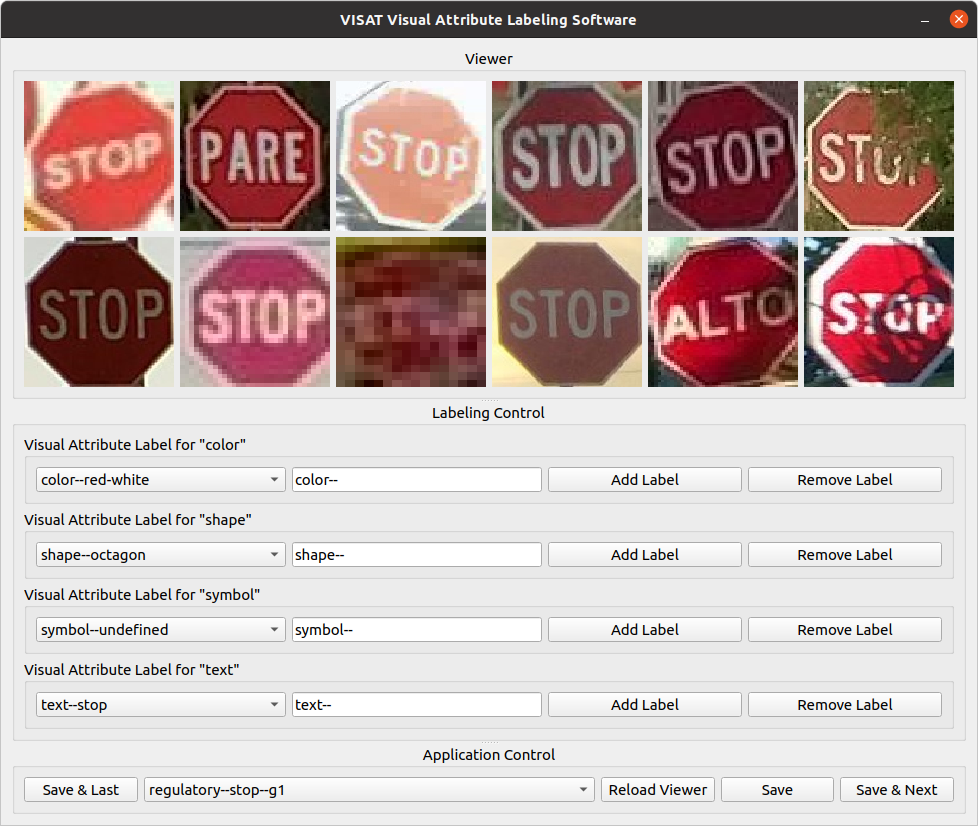}
        \caption{VISAT visual attribute labeling software.}
        \label{fig:labeling_software}
    \end{center}
    \vspace{-15pt}
\end{wrapfigure}

\paragraph{Rapid Visual Attribute Labeling Software} We developed a graphical software framework for the rapid creation of visual attributes, given the time-consuming nature of the task, even with only 401 classes.
As shown in Figure~\ref{fig:labeling_software}, the software includes 3 vertical divisions: Viewer, Labeling Control, and Application Control.
The Viewer randomly displays sign patches from a class and can be randomly refreshed by clicking the "Reload Viewer" button in the Application Control division, providing a comprehensive view of the fundamental features of the sign patches.
The number of displayed images is customizable to accommodate various screen sizes.
In the Labeling Control division, human labelers can create and select visual attributes for the current class.
New attributes are added by typing them into the text box and clicking "Add Label," while the drop-down menu enables attribute selection and removal.
The Application Control division enables saving of the attribute selections, sequential cycling through classes, and random access to any class.
Saved attribute selections are displayed when jumped to a class.
The software's Labeling Control division is fully configurable, accommodating any number or any type of user-defined attribute.
Therefore, our visual attribute labeling software is effectively dataset-agnostic and, in fact, can be used for creating visual attributes for any dataset where they can be conceptually identified and formulated by humans.
We intend to release this software to the public, and the information on the release can be found in Appendix~\ref{app:dataset}.

\begin{figure}
    \hspace{-8pt}
    \begin{minipage}{0.60\textwidth}
        \begin{center}
            \includegraphics[width=1.1\textwidth]{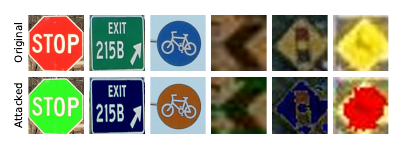}
            \caption{Color quantization examples.}
            \label{fig:color_quantization}
        \end{center}
    \end{minipage}
    \hspace{8mm}
    \begin{minipage}{0.35\textwidth}
        \begin{center}
            \vspace{5pt}
            \includegraphics[width=0.963\textwidth]{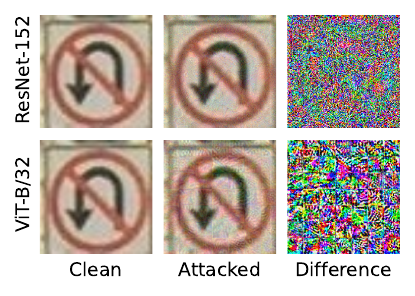}
            \vspace*{-4pt}
            \caption{PGD attack examples.}
            \label{fig:pgd_attack}
        \end{center}
    \end{minipage}
\end{figure}

\paragraph{Adversarial Attack} In order to evaluate the robustness of models against adversarial attacks, we applied the projected gradient descent (PGD) technique~\cite{madry2017towards} on the original VISAT testing split to generate adversarial examples specifically tailored for our dataset.
PGD is an iterative attack method widely utilized in adversarial robustness evaluation.
For all PGD attacks, we carefully selected a step size of 0.01 for 20 steps with $\epsilon=0.2$ and $\lVert \epsilon \rVert = \infty$.
By leveraging the gradients of the loss function, we iteratively crafted adversarial perturbations on each image in the dataset to maximize the vulnerability of the models, with examples shown in Figure~\ref{fig:pgd_attack}.
Notably, we designed two distinct adversarial testing splits: one for each of the models, namely ResNet-152, and ViT-B/32, used in this study.
Additionally, we extend our evaluation by generating 16 additional adversarial testing splits, where each of the 4 splits corresponds to the 4 MTL tasks in the MTL variant of the two models, respectively.
These adversarial examples, forming an integral part of the VISAT dataset, provide valuable resources for assessing model performance under various adversarial scenarios and gaining deeper insights into their susceptibility to adversarial attacks for traffic sign recognition.

\begin{figure}
    \begin{center}
        \includegraphics[width=\textwidth]{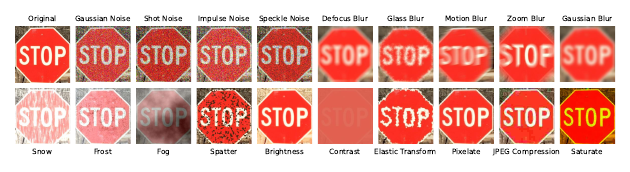}
        \vspace{-0.25in}
        \caption{Data corruption examples using ImageNet-C.}
        \label{fig:data_corruption}
    \end{center}
        \vspace{-3ex}
\end{figure}

\paragraph{Distribution Shift using Data Corruption} To comprehensively examine the impact of distribution shift on model robustness, we leverage the data corruption and variation techniques from ImageNet-C~\cite{hendrycks2019benchmarking}.
ImageNet-C offers a diverse range of 19 individual corruptions, encompassing 4 major variation types: noise, blur, weather, and digital.
These corruptions are characterized by varying levels of severity, spanning from mild (level 1) to severe (level 5).
To create a comprehensive evaluation framework, we utilize the codebase provided by ImageNet-C and generate a total of $19 \times 5 = 95$ sets of distribution shift testing splits from the original VISAT testing split, with examples from each corruption at severity 5 shown in Figure~\ref{fig:data_corruption}.
The resulting testing splits encompass all corruption types and severity levels, allowing us to thoroughly assess model performance under various distribution shifts.
By incorporating these varied and realistic corruption scenarios, we aim to provide a robust evaluation of model performance in the face of distribution shift challenges.

\paragraph{Distribution Shift using Color Quantization} We devise a color quantization technique to intentionally modify the color patterns of traffic signs in the original VISAT testing split, aiming to assess the impact of these alterations on the color MTL task as well as on the remaining tasks.
To perform color quantization, we utilized the k-means clustering algorithm to quantize elements of color on each image in the original testing split.
The resulting testing split allows us to delve into the intricate relationship amongst all the task heads in MTL models, shedding light on potential spurious correlations and uncovering the robustness of the MTL models under distribution shift.
Examples generated using color quantization can be seen in Figure~\ref{fig:color_quantization}.
The detailed discussions on color quantization, such as parameter selections and color swapping logic, can be found in Appendix~\ref{app:dataset}.

    \section{VISAT Robustness Benchmarks}
\label{sec:benchmarks}

\paragraph{Models} In our benchmarking experiments, we carefully selected two widely recognized vision backbones, ResNet-152 and ViT-B/32, as the base models for the benchmark evaluations of our VISAT dataset.
In the following benchmarks, the base ResNet-152 and ViT-B/32 models are denoted as ResNet-152-Base-Fine-Tuned (RBFT) and ViT-B/32-Base-Fine-Tuned (VBFT), respectively.
Expanding on the capabilities of our VISAT dataset, we extended our experiments to include the training of two additional MTL models, denoted as ResNet-152-MTL-Fine-Tuned (RMFT) and ViT-B/32-MTL-Fine-Tuned (VMFT).
These MTL models were trained using the visual attributes in our dataset, each of which encompasses 4 task heads, corresponding to the four visual attributes we created for VISAT.
Furthermore, to investigate the presence of spurious correlations among the MTL tasks, we trained two additional MTL models with their backbone fine-tuning disabled.
These models, denoted in the following benchmarks as ResNet-152-MTL-Fine-Tuned-Linear (RMFTL) and ViT-B/32-MTL-Fine-Tuned-Linear (VMFTL), shall provide insights into the impact of backbone fine-tuning on the performance and the level of spurious correlations within the MTL framework.
For all of our MTL models, we use the cross-entropy objective for each task head, and we weigh each task based on the size of its label space, i.e., the number of its corresponding type of visual attributes.
For i-th task, let us denote the size of its label space as $\vert S_i \vert$ and its model prediction and true label as $pred_i$ and $y_i$, respectively. Our overall MTL objective is then as follows:
\begin{equation*}
        \mathcal{L}_{\textit{MTL}} =  \displaystyle\sum_{i}  \frac{\mathcal{L}_{\textit{CE}}(pred_i, y_i)}{\log \vert S_i \vert}
\end{equation*}
The details on the training process of our models, together with the selection of hyperparameters, as well as information on the computational hardware used, can be found in Appendix~\ref{app:benchmarks}.

\paragraph{Metrics} For the following benchmarks, we use the metrics defined below to gauge model robustness:
\vspace{-3ex}
\begin{itemize}[leftmargin=*]
\setlength\itemsep{0.1em}
    \item $\epsilon$: model error, i.e., 1 - model accuracy.
    \item $\epsilon_{clean}$ or $\epsilon_{cl}$: model error evaluated on the original, clean VISAT testing split.
    \item $\epsilon_{attacked}$ or $\epsilon_{at}$: model error evaluated on an attacked VISAT testing split.
    \item $\Delta_\epsilon$= $\epsilon_{at}$ - $\epsilon_{cl}$: relative model error, i.e., difference between attacked and clean model errors.
    \item $\epsilon_{cumulative}$ or $\epsilon_{cu}$: cumulative relative model error across MTL tasks under an attack.
    \item $\textit{RECorr} = \frac{1}{n} \sum_{i = 0}^n \frac{\Delta_{\epsilon_i}}{\Delta_{\epsilon_t}}$: relative error correlation across MTL tasks under one attack targeting task $t$. $n$ is the number of non-targeted MTL tasks, in our case, 3. $\Delta_{\epsilon_\text{t}}$ is the relative model error of the targeted MTL task.
\end{itemize}
\begin{table}
    \begin{minipage}{0.48\textwidth}
        \caption{Base model error rates under PGD attacks.}
        \scriptsize
        \begin{center}
            \begin{tabular}{lllll}
                \toprule
                Models & Attacks & $\epsilon_\text{cl}$ & $\epsilon_\text{at}$ & $\Delta_\epsilon$ \\
                \midrule
                \multirow{2}{*}{RBFT}
                & RBFT & \multirow{2}{*}{7.50} & 99.96     & 92.46 \\
                & VBFT &                       & 99.96     & 92.46 \\
                \midrule
                \multirow{2}{*}{VBFT}
                & RBFT & \multirow{2}{*}{6.99} & 8.26      & 1.27  \\
                & VBFT &                       & 100.00    & 93.01 \\
                \bottomrule
            \end{tabular}
        \end{center}
        \label{tab:results_base_pgd}
    \end{minipage}
    \quad
    \begin{minipage}{0.48\textwidth}
        \caption{Base model error rates under ImageNet-C data corruption attacks.}
        \scriptsize
        \begin{center}
            \begin{tabular}{lllll}
                \toprule
                Models & Attacks & $\epsilon_\text{cl}$ & $\overline{\epsilon_\text{at}}$ & $\overline{\Delta_\epsilon}$ \\
                \midrule
                \multirow{4}{*}{RBFT}
                & Noise     & \multirow{4}{*}{7.50} & 36.26 & 28.76 \\
                & Blur      &                       & 8.50  & 1.00  \\
                & Weather   &                       & 30.13 & 22.64 \\
                & Digital   &                       & 13.81 & 6.31  \\
                \midrule
                \multirow{4}{*}{VBFT}
                & Noise     & \multirow{4}{*}{6.99} & 13.65 & 6.66  \\
                & Blur      &                       & 7.58  & 0.59  \\
                & Weather   &                       & 23.63 & 16.64 \\
                & Digital   &                       & 12.65 & 5.65  \\
                \bottomrule
            \end{tabular}
        \end{center}
        \label{tab:results_base_corruption}
    \end{minipage}
\end{table}

\begin{figure}
    \begin{minipage}{0.48\textwidth}
        \begin{center}
            \includegraphics[width=\textwidth]{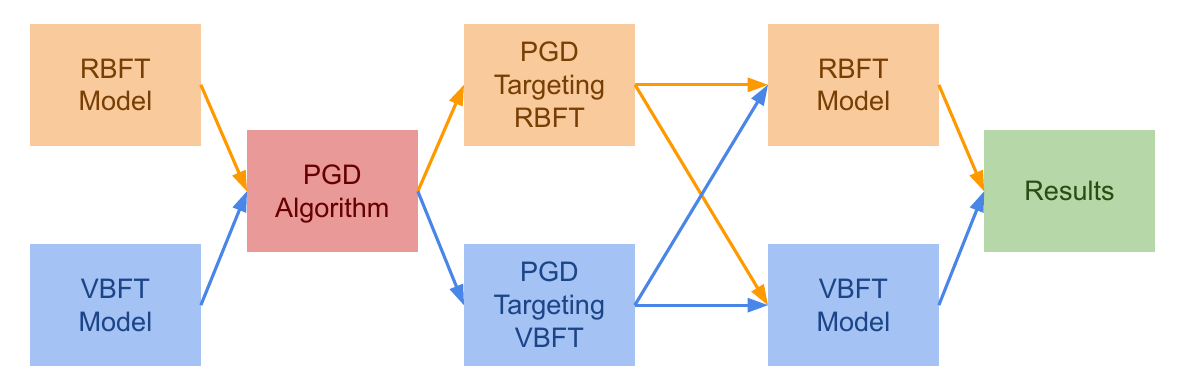}
            \caption{PGD attack schema for base models.}
            \label{fig:pgd_attack_schema_base}
        \end{center}
    \end{minipage}
    \quad
    \begin{minipage}{0.48\textwidth}
        \begin{center}
            \includegraphics[width=\textwidth]{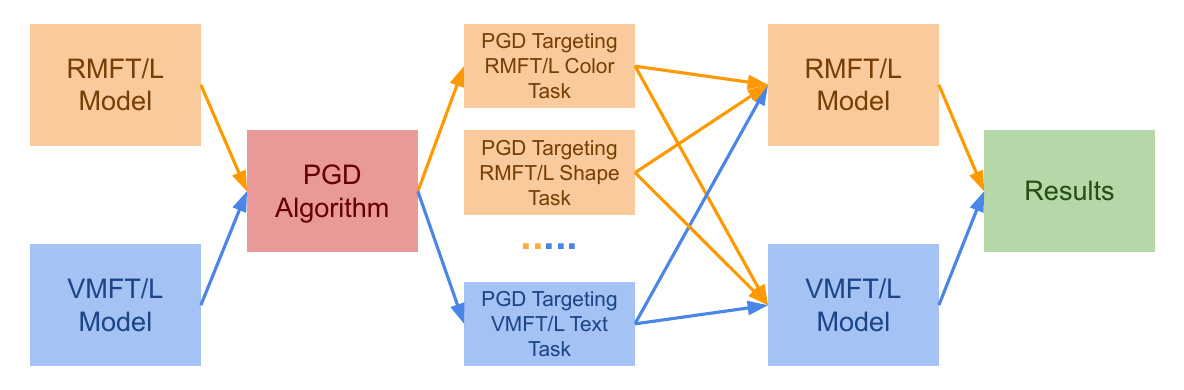}
            \caption{PGD attack schema for MTL models.}
            \label{fig:pgd_attack_schema_mtl}
        \end{center}
    \end{minipage}
\end{figure}

\paragraph{Adversarial Attack} As outlined in Section~\ref{sec:dataset}, we constructed two base adversarial testing splits specifically tailored for the RBFT and VBFT models.
Additionally, we generated adversarial testing splits for each of the 4 MTL tasks in the 4 MTL models, resulting in a total of 16 MTL adversarial testing splits.
In our initial experiment, we evaluate the performance of RBFT and VBFT on their respective PGD-attacked testing splits. Subsequently, following the attack scheme in Figure~\ref{fig:pgd_attack_schema_base}, we conduct cross-examinations by assessing the performance of RBFT and VBFT on each other's PGD-attacked testing splits.
Analyzing the results, we observe that the PGD algorithm is highly effective against its targeted models, substantially reducing their performance to nearly zero.
Interestingly, in the cross-examination scenario, we find that PGD attacks generated using RBFT are ineffective against VBFT by observing the $\Delta_\epsilon$ values in Table~\ref{tab:results_base_pgd}, while attacks generated using VBFT remain effective for RBFT. Performing cross-examinations on the effectiveness of attacks generated using one model on other models is meaningful. For example, certain malicious actors in real-world scenarios might deploy a comprehensive set of attacks generated from various publicly available models in order to overwhelm any targeted models, despite not knowing the target model details.

\begin{table}
    \caption{MTL model error rates and correlations under PGD attacks.}
    \tiny
    \begin{center}
        \begin{tabular}{p{0.03\textwidth}p{0.1\textwidth}p{0.02\textwidth}p{0.02\textwidth}p{0.03\textwidth}p{0.02\textwidth}p{0.02\textwidth}p{0.03\textwidth}p{0.02\textwidth}p{0.02\textwidth}p{0.03\textwidth}p{0.02\textwidth}p{0.02\textwidth}p{0.03\textwidth}ll}
            \toprule
            \multirow{3}{*}{Models} & \multirow{3}{*}{Attacks} & \multicolumn{3}{l}{Color} & \multicolumn{3}{l}{Shape} & \multicolumn{3}{l}{Symbol} & \multicolumn{3}{l}{Text} & \multirow{3}{*}{$\epsilon_\text{cu}$} & \multirow{3}{*}{\textit{RECorr}} \\
            \cmidrule{3-14}
            & & $\epsilon_\text{cl}$ & $\epsilon_\text{at}$ & $\Delta_\epsilon$ & $\epsilon_\text{cl}$ & $\epsilon_\text{at}$ & $\Delta_\epsilon$ & $\epsilon_\text{cl}$ & $\epsilon_\text{at}$ & $\Delta_\epsilon$ & $\epsilon_\text{cl}$ & $\epsilon_\text{at}$ & $\Delta_\epsilon$ & & \\
            \midrule
            \multirow{8}{*}{RMFT}
            & RMFT Color     &
            \multirow{8}{*}{8.61}   & 54.87 & 46.26 &
            \multirow{8}{*}{8.30}   & 53.66 & 45.36 &
            \multirow{8}{*}{11.33}  & 40.51 & 29.18 &
            \multirow{8}{*}{4.24}   & 15.95 & 11.71 &132.51 & 0.62\\
            & RMFT Shape     &       & 53.56 & 44.95 & & 55.74 & 47.44 && 40.76  &29.43 && 15.95 & 11.71& 133.53&0.60 \\
            & RMFT Symbol    &       & 48.54 & 39.93 & & 48.24 & 39.94 & & 43.79 &32.46 & &15.49 & 11.25&123.58 & 0.94\\
            & RMFT Text      &       & 44.48 & 35.87 & & 44.41 & 36.11 & & 37.96& 26.63& & 18.49& 14.25& 112.85& 2.31\\
            & VMFT Color     &       & 29.13 & 20.52 & & 28.06 & 19.76 & & 27.59&16.26 & &10.95 &6.71 &63.25 & 0.69\\
            & VMFT Shape     &       & 28.07 & 19.46 & & 28.47 & 20.17 & & 27.23& 15.90& &10.89 &6.65 & 62.19& 0.69\\
            & VMFT Symbol    &       & 27.48 & 18.87 & & 27.05 & 18.75 & & 28.69& 17.36& & 10.59&6.35 & 61.34& 0.84\\
            & VMFT Text      &       & 25.79 & 17.18 & & 25.28 & 16.98 & & 25.88& 14.55& & 12.16& 7.92&56.63 & 2.05\\
            \midrule
            \multirow{8}{*}{VMFT}
            & RMFT Color &
            \multirow{8}{*}{7.64}   & 10.16 &	2.52 &
            \multirow{8}{*}{7.60}   & 9.96	& 2.36 &
            \multirow{8}{*}{11.11}  & 13.10&	1.99 &
            \multirow{8}{*}{3.99}   & 4.98 &	0.99 & 7.86& 0.71\\
            & RMFT Shape & & 10.10	& 2.46 && 	10.04	& 2.44 &&	13.13&	2.02&&	4.94& 	0.95 &7.87 & 0.74\\
            & RMFT Symbol & & 9.78	& 2.14	 && 9.67 &	2.07 &&	13.47 &	2.36 &&	4.85 &	0.86 &7.43 & 0.71\\
            & RMFT Text & & 9.20	& 1.56 &&	9.02 &	1.42 &&	12.42 &	1.31 &&	5.08 &	1.09 &5.39 & 1.32\\
            & VMFT Color & & 99.98	& 92.34 &&	98.66 &	91.06 &&	88.17 &	77.06 &&	54.57 &	50.58
 &311.04 & 0.79 \\
            & VMFT Shape & &99.64	& 92.00 &&	99.99 &	92.39 &&	88.33 &	77.22&&	60.76 &	56.77 &318.38 &0.82 \\
            & VMFT Symbol & &92.64	 & 85.00 &&	93.25	&85.65	 && 99.93 &	88.82	 &&45.75	 & 41.76
 &301.23 & 0.80\\
            & VMFT Text & & 86.02	& 78.38 &&	89.20 &	81.60	&& 86.85 &	75.74 &&	99.83 &	95.84 &331.56 &0.82 \\
            \midrule
            \multirow{8}{*}{RMFTL}
            & RMFTL Color &
            \multirow{8}{*}{28.47} & 99.62 & 71.15 &
            \multirow{8}{*}{27.36} & 95.15	& 67.79 &
            \multirow{8}{*}{39.06} & 99.88 & 60.82 &
            \multirow{8}{*}{15.50} & 99.34 & 83.84 & 283.60 & 1.00 \\
            & RMFTL Shape  & & 99.62 & 71.15 & & 95.15 & 67.79 & & 99.88 & 60.82 & & 99.34 & 83.84 & 283.60 & 1.06\\
            & RMFTL Symbol & & 99.62 & 71.15 & & 95.15 & 67.79 & & 99.88 & 60.82 & & 99.34 & 83.84 & 283.60 & 1.22\\
            & RMFTL Text   & & 99.62 & 71.15 & & 95.15 & 67.79 & & 99.88 & 60.82 & & 99.34 & 83.84 & 283.60 & 0.79\\
            & VMFTL Color  & & 99.62 & 71.15 & & 95.15 & 67.79 & & 99.88 & 60.82 & & 99.34 & 83.84 & 283.60& 1.00\\
            & VMFTL Shape  & & 99.62 & 71.15 & & 95.15 & 67.79 & & 99.88 & 60.82 & & 99.34 & 83.84 & 283.60& 1.06\\
            & VMFTL Symbol & & 99.62 & 71.15 & & 95.15 & 67.79 & & 99.88 & 60.82 & & 99.34 & 83.84 & 283.60& 1.22\\
            & VMFTL Text   & & 99.62 & 71.15 & & 95.15 & 67.79 & & 99.88 & 60.82 & & 99.34 & 83.84 & 283.60& 0.79\\
            \midrule
            \multirow{8}{*}{VMFTL}
            & RMFTL Color &
            \multirow{8}{*}{19.92}   & 27.31 &	7.39 &
            \multirow{8}{*}{19.92}   & 24.67	& 4.75 &
            \multirow{8}{*}{31.88}  & 33.12 & 1.24 &
            \multirow{8}{*}{12.73}   & 13.51 &	0.78 &14.16 &0.31 \\
            & RMFTL Shape  & & 26.94  & 7.02  & & 25.16  &	5.24  & & 33.05  & 1.17  & & 13.65  & 0.92  & 14.35	 & 0.58 \\
            & RMFTL Symbol & & 25.76  & 5.84  & & 23.66  &	3.74  & & 33.18  & 1.30  & & 13.40  & 0.67  & 11.55  & 2.63 \\
            & RMFTL Text   & & 25.45  & 5.53  & & 23.27  &	3.35  & & 32.74  & 0.86  & & 13.64  & 0.91  & 10.65  & 3.55 \\
            & VMFTL Color  & & 100.00 & 80.08 & & 98.73  &	78.81 & & 88.77  & 56.89 & & 66.45  & 53.72 & 269.50 & 0.79 \\
            & VMFTL Shape  & & 99.23  & 79.31 & & 100.00 &	80.08 & & 88.41  & 56.53 & & 65.36  & 52.63 & 268.55 & 0.78 \\
            & VMFTL Symbol & & 97.80  & 77.88 & & 94.83  &	74.91 & & 100.00 & 68.12 & & 66.13  & 53.40 & 274.30 & 1.01 \\
            & VMFTL Text   & & 90.02  & 70.10 & & 94.96  &	75.04 & & 89.32  & 57.44 & & 100.00 & 87.27 & 289.85 & 0.77 \\
            \bottomrule
        \end{tabular}
        \vspace{-3ex}
    \end{center}
    \label{tab:results_mtl_pgd}
\end{table}

\begin{table}
    \caption{MTL model error rates under ImageNet-C data corruption attacks.}
    \scriptsize
    \begin{center}
        \begin{tabular}{p{0.04\textwidth}p{0.05\textwidth}p{0.03\textwidth}p{0.03\textwidth}p{0.04\textwidth}p{0.03\textwidth}p{0.03\textwidth}p{0.04\textwidth}p{0.03\textwidth}p{0.03\textwidth}p{0.04\textwidth}p{0.03\textwidth}p{0.03\textwidth}p{0.04\textwidth}l}
            \toprule
            \multirow{3}{*}{Models} & \multirow{3}{*}{Attacks} & \multicolumn{3}{l}{Color} & \multicolumn{3}{l}{Shape} & \multicolumn{3}{l}{Symbol} & \multicolumn{3}{l}{Text} & \multirow{3}{*}{$\epsilon_\text{cu}$} \\
            \cmidrule{3-14}
             & & $\epsilon_\text{cl}$ & $\overline{\epsilon_\text{at}}$ & $\overline{\Delta_\epsilon}$ & $\epsilon_\text{cl}$ & $\overline{\epsilon_\text{at}}$ & $\overline{\Delta_\epsilon}$ & $\epsilon_\text{cl}$ & $\overline{\epsilon_\text{at}}$ & $\overline{\Delta_\epsilon}$ & $\epsilon_\text{cl}$ & $\overline{\epsilon_\text{at}}$ & $\overline{\Delta_\epsilon}$ & \\
            \midrule
            \multirow{4}{*}{RMFT}
            & Noise & \multirow{4}{*}{8.61}  & 43.22 & 34.62 &
                      \multirow{4}{*}{8.30}  & 42.60 & 34.31 &
                      \multirow{4}{*}{11.33} & 35.36 & 24.03 &
                      \multirow{4}{*}{4.24}  & 14.12 & 9.87  & 102.82 \\
            & Blur &                         & 9.75  & 1.15  &
                                             & 9.61  & 1.32  &
                                             & 12.73 & 1.40  &
                                             & 5.18  & 0.94  & 4.80 \\
            & Weather &                      & 32.39 & 23.79 &
                                             & 32.03 & 23.74 &
                                             & 29.91 & 18.58 &
                                             & 11.90 & 7.66  & 73.77 \\
            & Digital &                      & 14.79 & 6.19  &
                                             & 14.24 & 5.94  &
                                             & 17.33 & 6.00  &
                                             & 7.07  & 2.83  & 20.96 \\
            \midrule
            \multirow{4}{*}{VMFT}
            & Noise & \multirow{4}{*}{7.64}  & 17.33 & 9.68  &
                      \multirow{4}{*}{7.60}  & 16.75 & 9.15  &
                      \multirow{4}{*}{11.11} & 18.60 & 7.49  &
                      \multirow{4}{*}{3.99}  & 7.97  & 3.98  & 30.30 \\
            & Blur &                         & 8.25  & 0.61  &
                                             & 8.23  & 0.63  &
                                             & 11.70 & 0.59  &
                                             & 4.48  & 0.48  & 2.31  \\
            & Weather &                      & 25.70 & 18.05 &
                                             & 25.27 & 17.67 &
                                             & 24.91 & 13.81 &
                                             & 10.08 & 6.08  & 55.62 \\
            & Digital &                      & 13.65 & 6.01  &
                                             & 13.25 & 5.65  &
                                             & 15.48 & 4.37  &
                                             & 6.11  & 2.12  & 18.15 \\
            \midrule
            \multirow{4}{*}{RMFTL}
            & Noise & \multirow{4}{*}{28.47} & 51.09 & 22.62 &
                      \multirow{4}{*}{27.36} & 50.97 & 23.61 &
                      \multirow{4}{*}{39.06} & 39.23 & 0.17  &
                      \multirow{4}{*}{15.50} & 15.57 & 0.07  & 46.47 \\
            & Blur &                         & 33.47 & 5.00  &
                                             & 33.03 & 5.68  &
                                             & 39.22 & 0.16  &
                                             & 15.54 & 0.05  & 10.88 \\
            & Weather &                      & 46.25 & 17.79 &
                                             & 44.76 & 17.40 &
                                             & 39.23 & 0.17  &
                                             & 15.56 & 0.07  & 35.43 \\
            & Digital &                      & 38.06 & 9.59  &
                                             & 36.33 & 8.97  &
                                             & 39.22 & 0.16  &
                                             & 15.55 & 0.06  & 18.78 \\
            \midrule
            \multirow{4}{*}{VMFTL}
            & Noise & \multirow{4}{*}{19.92} & 30.08 & 10.16 &
                      \multirow{4}{*}{19.12} & 28.03 & 8.91  &
                      \multirow{4}{*}{31.88} & 34.39 & 2.51  &
                      \multirow{4}{*}{12.73} & 14.15 & 1.41  & 23.00 \\
            & Blur &                         & 27.74 & 7.82  &
                                             & 27.20 & 8.08  &
                                             & 34.19 & 2.31  &
                                             & 13.90 & 1.16  & 19.37 \\
            & Weather &                      & 31.10 & 11.19 &
                                             & 30.42 & 11.31 &
                                             & 34.83 & 2.95  &
                                             & 14.17 & 1.44  & 26.88 \\
            & Digital &                      & 25.94 & 6.02  &
                                             & 24.21 & 5.09  &
                                             & 34.25 & 2.37  &
                                             & 13.76 & 1.03  & 14.51 \\
            \bottomrule
        \end{tabular}
        
        \vspace{-3ex}
    \end{center}
    \label{tab:results_mtl_corruption}
\end{table}

\begin{table}
    \caption{MTL model error rates and correlations under color quantization attacks.}
    \scriptsize
    \begin{center}
        \begin{tabular}{p{0.05\textwidth}p{0.03\textwidth}p{0.03\textwidth}p{0.04\textwidth}p{0.03\textwidth}p{0.03\textwidth}p{0.04\textwidth}p{0.03\textwidth}p{0.03\textwidth}p{0.04\textwidth}p{0.03\textwidth}p{0.03\textwidth}p{0.04\textwidth}p{0.03\textwidth}p{0.05\textwidth}}
            \toprule
            \multirow{3}{*}{Models} & \multicolumn{3}{l}{Color} & \multicolumn{3}{l}{Shape} & \multicolumn{3}{l}{Symbol} & \multicolumn{3}{l}{Text} & \multirow{3}{*}{$\epsilon_\text{cu}$} & \multirow{3}{*}{\textit{RECorr}}\\
            \cmidrule{2-13}
             & $\epsilon_\text{cl}$ & $\epsilon_\text{at}$ & $\Delta_\epsilon$ & $\epsilon_\text{cl}$ & $\epsilon_\text{at}$ & $\Delta_\epsilon$ & $\epsilon_\text{cl}$ & $\epsilon_\text{at}$ & $\Delta_\epsilon$ & $\epsilon_\text{cl}$ & $\epsilon_\text{at}$ & $\Delta_\epsilon$ & & \\
            \midrule
            RMFT & 8.61 & 32.22 & 23.61 &
                  8.30 & 25.91 & 17.61 &
                  11.33 & 26.54 & 15.21 &
                  4.24 & 12.45 & 8.21 &
                  64.64 & 0.58 \\
            \midrule
            VMFT & 7.64 & 31.90 & 24.26 &
                  7.60 & 26.54 & 18.94 &
                  11.11 & 26.42 & 15.32 &
                  3.99 & 12.21 & 8.22 &
                  66.74 & 0.58 \\
            \midrule
            RMFTL & 28.47 & 46.99 & 18.53 &
                   27.36 & 37.45 & 10.10 &
                   39.06 & 39.22 & 0.16 &
                   15.50 & 15.57 & 0.07 &
                   28.85 & 0.19 \\
            \midrule
            VMFTL & 19.92 & 36.37 & 16.45 &
                   19.12 & 27.29 & 8.18 &
                   31.88 & 35.81 & 3.93 &
                   12.73 & 14.72 & 1.99 &
                   30.55 & 0.29 \\
            \bottomrule
        \end{tabular}
    \end{center}
    \label{tab:results_mtl_color_quantization}
\end{table}

Moving forward, we evaluate the performance of RMFT/L and VMFT/L models.
Similar to the base evaluations, we follow the attack scheme illustrated in Figure~\ref{fig:pgd_attack_schema_mtl} and cross-examine the performance of the 4 MTL models across the 16 adversarial testing splits.
We observe by examining the values of $\epsilon_{cu}$ in Table~\ref{tab:results_mtl_pgd} that attacks generated using VMFT/L effectively compromise RMFT/L, while attacks generated using RMFT/L are ineffective against VMFT/L.
Additionally, we find that the targeted tasks consistently exhibit the largest $\Delta_\epsilon$.
Moreover, through cross-examination, we uncover the presence of noticeable \textit{RECorr} among the tasks.
Attacks targeting one specific task result in noticeable $\Delta_\epsilon$ in other tasks, providing compelling signs of error correlations among MTL tasks.
Interestingly, we further observe that the error rates of RMFTL show similarity across RMFTL and VMFTL attacks on all tasks, with minute variations invisible in Table~\ref{tab:results_mtl_pgd}.
Such consistent artifact may be attributed to the fact that the PGD attacks solely target the task heads with backbone fine-tuning disabled, causing attacks on different tasks to have virtually the same effect across all tasks for ResNet-152.

\paragraph{Distribution Shift} To comprehensively assess model robustness under distribution shift, we utilized 95 testing splits generated by applying various corruptions and severity levels from the ImageNet-C dataset, previously described in Section~\ref{sec:dataset}.
For results in Table~\ref{tab:results_base_corruption} and~\ref{tab:results_mtl_corruption}, we averaged $\epsilon_{at}$ and $\Delta_\epsilon$ across the 4 types of variations, noise, blur, weather, and digital. Tables containing complete and detailed results across all individual 95 corruption testing splits can be found in Appendix~\ref{app:benchmarks}.
First, we evaluated the performance of RBFT and VBFT models on these testing splits.
The results in Table~\ref{tab:results_base_corruption} reveal that, overall, VBFT exhibits greater robustness compared to RBFT, with a relatively smaller $\overline{\Delta_\epsilon}$.
Furthermore, we observe corruptions that specifically target image resolution, such as the blur type, are relatively ineffective for both models with the lowest $\overline{\Delta_\epsilon}$ values in most cases.
This observation aligns with the characteristics of our dataset, where sign patches cropped from original MTSD full-frame images, often appearing at a distance, are, therefore, small in size.
Consequently, the blurring effects are already present in the training and validation datasets, rendering them ineffective as distribution shifts during testing.
Conversely, weather-related attacks cause noticeable reductions in the performance of both models with relatively high $\overline{\Delta_\epsilon}$ values, demonstrating their effectiveness. 

Next, we evaluate the performance of RMFT and VMFT on the corruption testing splits.
By inspecting the respective $\epsilon_{cu}$ values, we observe the results shown in Table~\ref{tab:results_mtl_corruption} exhibiting a similar trend to the base evaluations, with blurring corruptions remaining ineffective while weather-based attacks continue to impact the models.
Moreover, VMFT demonstrates greater robustness overall compared to RMFT, with relatively smaller $\epsilon_{cu}$ values.
Additionally, we find that the color and shape tasks are the least robust against the attacks, experiencing significant $\overline{\Delta_\epsilon}$ across the board. Conversely, the text task exhibits the highest level of robustness against all attacks.
We also evaluated the performances of RMFTL and VMFTL and saw noticeably worse overall robustness in both models with large increases in $\epsilon_{cu}$ values due to the lack of backbone fine-tuning.
Moving on to the evaluation of our color quantization attacks, we applied the color quantization testing split to all 4 MTL models.
As an out-of-distribution attack specifically targeting the color task, minimal to no impact on the other 3 tasks would be expected if there were little to no spurious correlations among the tasks.
However, the results reveal noticeable \textit{RECorr} for all 4 models under attacks specifically targeting their color task.

The error bars and the reproducibility of all the above benchmark results, together with details on the release of our codebase developed for training and evaluations, are in Appendix~\ref{app:benchmarks}.
\vspace{-1ex}

    \section{Discussion and Conclusion}
\label{sec:conclusion}
\vspace{-1ex}

In this paper, we present VISAT, a comprehensive dataset and robustness benchmark suite for traffic sign recognition with visual attributes for both single-task and multi-task learning settings.
Yet, it is crucial for us to address the limitations and potential ethical concerns of our work.
\vspace{-1ex}

\paragraph{Limitations and Future Work} Our exploration of model robustness primarily focuses on classification prediction, leaving room for future work to include robustness assessments for bounding box predictions involving models such as R-CNNs~\cite{girshick2014rich} and YOLOs~\cite{redmon2016you}.
Additionally, our color quantization attacks have limited performance on low-resolution images, as seen in Figure~\ref{fig:color_quantization}.
Nevertheless, the approach adequately demonstrates spurious correlations within MTL models.
Future research, however, should expand attacks beyond color to target shape, symbol, and text tasks for a comprehensive assessment of spurious correlations across various MTL tasks.  
Furthermore, exploring alternative adversarial algorithms, enhancing the dataset with combined natural variations to better represent real-world scenarios, and creating visual attributes at the instance level for more accurate labeling are areas that warrant further investigation.
\vspace{-1ex}

\paragraph{Safety and Ethics}
While we have made significant efforts to ensure the quality and relevance of our work, it is crucial to acknowledge potential ethical implications and safety concerns that arise from the use of our dataset and benchmarks.
Our dataset, based on MTSD~\cite{mapillary2022mapillary}, comprises street-captured full-frame images, which may include the appearances of individuals or personal vehicles. However, we strictly slice the traffic signs from full-frame images using their bounding box annotations, effectively eliminating any personally identifiable information.
Moreover, researchers and engineers must ensure that any deployment or application of models trained using our dataset adheres to rigorous safety standards, legal requirements, and ethical guidelines.
Thorough testing, validation, and assessments are essential before deploying such models in real-world scenarios.

In conclusion, the creation of a rich set of visual attributes in our dataset provides researchers with a valuable resource for exploring spurious correlations and vulnerabilities within MTL models.
Our benchmarks facilitate the evaluation of model performance under adversarial attacks and distribution shifts, shedding light on potential robustness threats.
We acknowledge the need for further research to address the identified limitations, expand the range of attacks, and experiment with a broader set of models.
By highlighting these aspects, we hope that our dataset and benchmarks will inspire future studies, promoting the development of more robust and reliable machine learning models for real-world applications while fostering responsible and ethical ML practices.

    \bibliographystyle{plain}

    \newpage
\appendix

\section{Extended Related Work, Scopes, and Rationales}
\label{app:related_work}

\subsection{Traffic Sign Recognition}

\paragraph{Models} Image recognition and classification models have seen significant advancements, particularly with the evolution of convolutional neural network (CNN) architectures.
Ever since the conception of AlexNet~\cite{krizhevsky2017imagenet}, models such as VGG~\cite{simonyan2014very}, GoogLeNet~\cite{szegedy2015going, szegedy2016rethinking}, ResNet~\cite{he2016deep, he2016identity}, DenseNet~\cite{huang2017densely}, and Vision Transformer (ViT)~\cite{dosovitskiy2020image} have demonstrated the benefits of increasing network depth and width, introducing shortcut connections, exploring alternative layer connections, and reshaping images into flattened patches for transformer encoders.
In addition to classification predictions, there exists a distinct line of models that also produce bounding box predictions, including R-CNNs~\cite{girshick2014rich, girshick2015fast, ren2015faster, he2017mask}, YOLOs~\cite{redmon2016you, redmon2017yolo9000, redmon2018yolov3, bochkovskiy2020yolov4, wang2021scaled}, and RetinaNet~\cite{lin2017focal}.

In this work, we primarily focus on classification robustness and employ popular image classification models, specifically, ResNet-152 and ViT-B/32, for our baseline evaluations.

\subsection{Multi-Task Learning and Ensembles}

\paragraph{Domain Knowledge Integration}

Multi-task learning utilizes domain information from related tasks to improve generalization~\cite{caruana1997multitask}.
Datasets like BDD100K~\cite{yu2020bdd100k}, MEDIC~\cite{alam2023medic}, and MLM~\cite{armitage2020mlm} support MTL research in evaluating image recognition algorithms, disaster image classification, and multi-task learning with multiple languages and modalities, respectively.
In traffic sign recognition, a multi-modal tree-structure embedded MTL approach~\cite{lu2016traffic} integrates different visual features and modalities for enhanced sign recognition.
Furthermore, several studies~\cite{gurel2021knowledge, zhang2022care, yang2022improving} highlight the growing interest in leveraging additional attributes to train and evaluate ensemble networks, emphasizing the significance of domain-specific knowledge and attribute-based approaches to enhance robustness.
Visual attributes improve robustness by considering diverse semantic tasks, while integrating first-order logic and probabilistic graphical models provides a principled methodology for combining ensemble network outputs and achieving certified robustness against adversarial attacks.

In line with this growing body of knowledge, our work makes a notable contribution by introducing a detailed set of visual attributes tailored for traffic sign recognition.
These attributes offer insights into the distinct characteristics and visual cues in traffic sign images.
By incorporating them into our dataset and benchmarks, researchers can readily adopt and integrate visual attribute-based approaches in their MTL or ensemble frameworks, expanding the application of visual attributes in conjunction with MTL or ensemble techniques to enhance model robustness.

\paragraph{Spurious Correlations}

Several studies have highlighted the presence and impact of spurious correlations in machine learning (ML) models. One work focused on recognition algorithms and their poor generalization to new environments~\cite{beery2018recognition}, demonstrating the effect of spurious correlation through the example of recognizing cows in different backgrounds. Another study explored the challenges of spurious correlation in MTL models specifically~\cite{hu2022improving}, showing that non-causal knowledge and label-label confounders can negatively affect generalization. Additionally, an empirical study on robustness to spurious correlations in NLP models~\cite{tu2020empirical} revealed the reliance on spurious correlations due to limited linguistic variations in training data. Although not directly related to our application, the work demonstrates the prevalence of spurious correlation as a common problem across domains.

These findings motivate our experiments on spurious correlations in traffic sign recognition, as we aim to investigate the robustness of MTL models to such correlations and motivate our communities to develop approaches against our benchmarks to mitigate their impact.

    \newpage
\section{Extended Details on VISAT Dataset}
\label{app:dataset}

\subsection{Public Access}
\label{app:dataset_access}

We release the VISAT dataset and benchmarks to the public via the official VISAT website~\footnote{VISAT website: \url{http://rtsl-edge.cs.illinois.edu/visat/}}.
The structured metadata describing the VISAT dataset is embedded into the index page of the VISAT website.
The download access of the VISAT dataset and its associated codebase, together with documentation and instructions, can be found on the downloads page of the VISAT website~\footnote{VISAT downloads page: \url{http://rtsl-edge.cs.illinois.edu/visat/downloads/}}.

\subsection{Explicit Licensing}
\label{app:dataset_licensing}

The Mapillary Traffic Sign Dataset (MTSD)~\cite{ertler2020mapillary, mapillary2022mapillary}, upon which our dataset is built, is released under the Creative Commons Attribution NonCommercial ShareAlike (CC BY-NC-SA) license.
Therefore, due to the ShareAlike term of the license, we are required to release our VISAT dataset and benchmarks under the same license, i.e., the Creative Commons Attribution NonCommercial ShareAlike (CC BY-NC-SA) license, as that of the MTSD.
A copy of the license for the VISAT dataset and benchmarks can be accessed at the VISAT website.
Prior to accessing our dataset and benchmarks, all users must read, understand, and follow the terms of the license~\footnote{VISAT license: \url{https://creativecommons.org/licenses/by-nc-sa/4.0/}}.

\subsection{Statement of Responsibility}
\label{app:dataset_responsibility}

We, the authors of the VISAT dataset and benchmarks, bear all responsibility in case of violation of rights due to the development and release of the VISAT project.
To once again confirm, as stated above, the VISAT dataset and benchmarks are released to the public under the Creative Commons Attribution NonCommercial ShareAlike (CC BY-NC-SA) license.

\subsection{Maintenance Plan}
\label{app:dataset_maintenance}

The release of the VISAT dataset and benchmarks is hosted on one of our laboratory servers within the University of Illinois network.
The VISAT project website hosted by our server is entirely open and uncredentialized.
The University of Illinois Technology Services will ensure the continuous functioning of our laboratory server hardware, while our project members will maintain all software on our server required by the VISAT project website and its release.
Therefore, the release of the VISAT dataset and benchmarks shall remain available to the public indefinitely.
If, for any reason, we elect to relocate the release of the VISAT dataset and benchmarks to other hosting websites or platforms, we will maintain public access to the original website of the release, where we would then indicate the location and access of the new hosting platform.

\subsection{Visual Attributes}

Upon completing the visual attribute creation process, our visual attribute labeling software generates two sets of information: the created visual attribute labels (color, shape, symbol, and text) and the mapping between the MTSD class labels and their corresponding attribute labels.
These two sets of information enable the creation of a custom data loader specifically designed for MTL models.
During training, validation, and testing, the custom MTL data loader provides the visual attribute labels as the ground truth for each of the four MTL task heads for every input image.
The size of the output layer for each MTL task head corresponds to the number of attribute labels in the respective visual attribute type.
For instance, if there are a total of 20 color attribute labels, then the output layer size for the color MTL task head is 20.
The created visual attribute labels and their statistics are listed on the VISAT website and are also available as JSON files in the dataset downloads.
The release of our rapid visual attribute labeling software, together with its documentation and instructions, can be found on the VISAT downloads page.

\subsection{Color Quantization}

We employ the k-means clustering algorithm to quantize or cluster elements of color in each traffic sign image.
The k-means algorithm involves three key parameters: the total number of clusters (k), the number of attempts, and the maximum number of iterations.
To determine the appropriate number of clusters, we manually varied k from 1 to 5 while keeping the other two parameters fixed and evaluated their individual performance.
Given the relatively small number of clusters we have, we set the number of attempts and maximum iterations to 100 to ensure sufficient explorations of initial RGB values and as many iterations as computationally allowed.
Prior to clustering, we vectorized the images to enhance clustering efficiency.
After careful evaluation, we selected k to be 3 and then identified the largest cluster with the lowest variance in distances from all pixels within that cluster to the image center.
Next, we examined the average RGB value of the selected cluster and randomly swapped the largest RGB value with one of the other two colors if the color channel difference between the largest and second largest values exceeded or equaled a predefined threshold (35).
In cases where the background color of the traffic sign was either white or black, indicating a small difference among the RGB values, we randomly set two out of the three RGB channels to 0, effectively rendering the color of the cluster to either red, green, or blue.

    \newpage
\section{Extended Details and Results on VISAT Robustness Benchmarks}
\label{app:benchmarks}

\begin{table}
    \caption{Hyperparameters used during model training.}
    \scriptsize
    \begin{center}
        \begin{tabular}{lll}
            \toprule
            Hyperparameters                 & ResNet-152    & ViT-B/32  \\
            \midrule
            LRS Cooldown                    & 0             & 0         \\
            LRS Factor                      & 0.1           & 0.1       \\
            LRS Minimum Learning Rate       & 0             & 0         \\
            LRS Minimum Learning Rate Decay & 1e-8          & 1e-8      \\
            LRS Mode                        & min           & min       \\
            LRS Patience                    & 1             & 1         \\
            LRS Threshold                   & 1e-4          & 1e-4      \\
            LRS Threshold Mode              & rel           & rel       \\
            Optimizer Learning Rate         & 1e-3          & 1e-3      \\
            Optimizer Momentum              & 0.9           & 0.9       \\
            Optimizer Weight Decay          & 1e-6          & 1e-6      \\
            \bottomrule
        \end{tabular}
    \end{center}
    \label{tab:training_hyperparameters}
\end{table}

\begin{table}
    \caption{Additional settings for model training.}
    \scriptsize
    \begin{center}
        \begin{tabular}{lll}
            \toprule
            Settings            & ResNet-152        & ViT-B/32          \\
            \midrule
            Batch Size          & 512               & 384               \\
            Epoch (Base)        & 50                & 50                \\
            Epoch (MTL)         & 100               & 100               \\
            Input Height        & 224               & 224               \\
            Input Width         & 224               & 224               \\
            Pretrained Weights  & IMAGENET1K\_V2    & IMAGENET1K\_V1    \\
            \bottomrule
        \end{tabular}
    \end{center}
    \label{tab:training_settings}
\end{table}

\subsection{Model Training}
\label{app:benchmarks_training}

The hyperparameters used during model training are listed under Table~\ref{tab:training_hyperparameters}, where LRS stands for the learning rate scheduler.
We used the same set of hyperparameters for the training of all models, i.e., ResNet-152, ViT-B/32, and all of their variants.
Additional settings we used during training are also listed under Table~\ref{tab:training_settings}.
Notably, we use a batch size of 512 for ResNet-152 and its variants and 384 for ViT-B/32 and its variants.
The losses of all base models converge well within 50 epochs, while those of all MTL models require 100 epochs to fully converge.

\begin{table}[H]
    \caption{Computational resources used for the VISAT project.}
    \scriptsize
    \begin{center}
        \begin{tabular}{llll}
            \toprule
            Name                            & CPU                           & GPU                               & RAM       \\
            \midrule
            Alienware Aurora R11            & Intel Core i9-10900K          & 2 x NVIDIA GeForce RTX 2070 Super & 32 GB     \\
            Brown Virtual Computing Nodes   & Intel Xeon Gold 6126          & 4 x NVIDIA Tesla V100             & 96 GB     \\
            Euler Node                      & Intel Xeon Silver 4314        & 4 x NVIDIA RTX A6000              & 256 GB    \\
            Fractal Define R5               & Intel Xeon E5-2682 v4         & 1 x NVIDIA GeForce RTX 3080       & 32 GB     \\
            Fractal Define XL R2            & AMD Ryzen Threadripper 2950X  & 2 x NVIDIA GeForce RTX 2080 Ti    & 64 GB     \\
            \bottomrule
        \end{tabular}
    \end{center}
    \label{tab:computational_resources}
\end{table}

\subsection{Computational Resources}
\label{app:benchmarks_resources}

Given the large size of the models and the scale of our dataset, substantial computational resources were required.
We leveraged multiple computational clusters available to us for efficient training and testing of the models as listed in Table~\ref{tab:computational_resources}.
To maximize the efficiency of our evaluations, we pipelined our experiments by performing model training on powerful hardware, such as the Euler Node and Fractal Define XL R2, while, in the meantime, performing inference and benchmarks of the models on smaller hardware, such as the Alienware Aurora R11 and Fractal Define R5.
The PGD attacks are generated on the Euler and Brown Virtual Computing Nodes.
The choices of batch sizes in Table~\ref{tab:training_settings} are also determined by the total GPU memory available on the Euler Node and Fractal Define XL R2.

\subsection{Reproducibility}
\label{app:benchmarks_reproducibility}

During the development and testing process of the VISAT dataset and benchmarks, we ensure the reproducibility of our dataset and results by taking the following measures:
\begin{itemize}
    \item We document all hyperparameters and additional settings used for model training, as listed in Table~\ref{tab:training_hyperparameters} and~\ref{tab:training_settings}.
    \item We perform all dataset generations and benchmark evaluations under a fixed set of random seeds. For our error bar results, we used seeds 42, 84, 126, 168, and 210.
    \item We ensure all experiments are performed with the same library versions across all the computational platforms.
    \item We release our dataset and benchmark codebase to the public with the necessary documentation for others to reproduce our results.
\end{itemize}

The release of our codebase used for model training and benchmark evaluations, together with its documentation and instructions, can be found on the VISAT downloads page~\footnote{VISAT downloads page: \url{http://rtsl-edge.cs.illinois.edu/visat/downloads/}}.

\begin{table}
    \caption{Base model error rates under PGD attacks with error bars.}
    \scriptsize
    \begin{center}
        \begin{tabular}{llllllll}
            \toprule
            Models & Attacks & \multicolumn{2}{l}{$\epsilon_\text{cl}$} & \multicolumn{2}{l}{$\epsilon_\text{at}$} & \multicolumn{2}{l}{$\Delta_\epsilon$}  \\
            \midrule
            \multirow{2}{*}{RBFT} & RBFT & \multirow{2}{*}{7.6923} & \multirow{2}{*}{$\pm 0.1217$} & 90.1602 & $\pm 21.7534$ & 82.4679 & $\pm 21.6460$ \\
            & VBFT & & & 90.1615 & $\pm 21.7541$ & 82.4692 & $\pm 21.6467$ \\
            \midrule
            \multirow{2}{*}{VBFT} & RBFT & \multirow{2}{*}{6.9602} & \multirow{2}{*}{$\pm 0.0503$} & 87.3579 & $\pm 6.6646$ & 79.6797 & $\pm 7.2312$ \\
            & VBFT & & & 85.0979 & $\pm 6.4383$ & 77.4197 & $\pm 6.9271$ \\
            \bottomrule
        \end{tabular}
    \end{center}
    \label{tab:results_complete_base_pgd}
\end{table}

\subsection{Extended Results}

The extended results include error bars, i.e., standard deviations, around the average values across 5 trials.
Each of the 5 trials uses seeds 42, 84, 126, 168, and 210, respectively.
Additionally, $\overline{\epsilon_{at}}$ and $\overline{\Delta_\epsilon}$ in Table~\ref{tab:results_base_corruption} and~\ref{tab:results_mtl_corruption} are extended into individual values for all 95 ImageNet-C data corruption attacks. 
Results in Section~\ref{sec:benchmarks} are values from the trial using seed 42.

Table~\ref{tab:results_base_pgd} is extended into Table~\ref{tab:results_complete_base_pgd}.
In the following pages, Table~\ref{tab:results_base_corruption} is extended into Table~\ref{tab:results_complete_base_corruption_rbft_1}-~\ref{tab:results_complete_base_corruption_vbft_2}.
Table~\ref{tab:results_mtl_pgd} is extended into Table~\ref{tab:results_complete_mtl_pgd_color}-~\ref{tab:results_complete_mtl_pgd_cumulative}.
Table~\ref{tab:results_mtl_corruption} is extended into Table~\ref{tab:results_complete_mtl_corruption_rmft_color_1}-~\ref{tab:results_complete_mtl_corruption_vmftl_cumulative_2}.
Table~\ref{tab:results_mtl_color_quantization} is extended into Table~\ref{tab:results_complete_mtl_color_quantization_color}-~\ref{tab:results_complete_mtl_color_quantization_cumulative}.

\newpage

\begin{table}
    \begin{minipage}{0.49\textwidth}
        \caption{RBFT error rates under individual ImageNet-C data corruption attacks (1-45) with error bars.}
        \tiny
        \begin{center}
\begin{tabular}{p{0.08\textwidth}p{0.06\textwidth}p{0.1\textwidth}p{0.04\textwidth}p{0.11\textwidth}p{0.04\textwidth}p{0.13\textwidth}}
                \toprule
                Attacks & Severity & $\epsilon_\text{cl}$ & \multicolumn{2}{l}{$\epsilon_\text{at}$} & \multicolumn{2}{l}{$\Delta_\epsilon$} \\
                \midrule
                \multirow{5}{10pt}{Gaussian Noise} & 1 & \multirow{5}{10pt}{7.6914 $\pm 0.1103$} & 16.9085 & $\pm 2.4382$ & 9.2171 & $\pm 2.4028$ \\
                & 2 & & 26.9121 & $\pm 4.1891$ & 19.2207 & $\pm 4.1711$ \\
                & 3 & & 39.5919 & $\pm 4.1740$ & 31.9004 & $\pm 4.1683$ \\
                & 4 & & 47.7667 & $\pm 2.0984$ & 40.0753 & $\pm 2.1049$ \\
                & 5 & & 50.8345 & $\pm 0.3756$ & 43.1430 & $\pm 0.4089$ \\
                \midrule
                \multirow{5}{10pt}{Shot Noise} & 1 & \multirow{5}{10pt}{7.6914 $\pm 0.1103$} & 16.9798 & $\pm 1.7412$ & 9.2884 & $\pm 1.7113$ \\
                & 2 & & 27.6710 & $\pm 3.9752$ & 19.9795 & $\pm 3.9571$ \\
\bottomrule
            \end{tabular}
        \end{center}
        \label{tab:results_complete_base_corruption_rbft_1}
    \end{minipage}
    \hspace{2pt}
    \begin{minipage}{0.49\textwidth}
        \caption{RBFT error rates under individual ImageNet-C data corruption attacks (46-95) with error bars.}
        \tiny
        \begin{center}
            \begin{tabular}{p{0.08\textwidth}p{0.06\textwidth}p{0.1\textwidth}p{0.04\textwidth}p{0.11\textwidth}p{0.04\textwidth}p{0.13\textwidth}}
                \toprule
                Attacks & Severity & $\epsilon_\text{cl}$ & \multicolumn{2}{l}{$\epsilon_\text{at}$} & \multicolumn{2}{l}{$\Delta_\epsilon$} \\
                \midrule
                \multirow{5}{10pt}{Snow} & 1 & \multirow{5}{10pt}{7.6914 $\pm 0.1103$} & 28.3611 & $\pm 1.7687$ & 20.6697 & $\pm 1.7185$ \\
                & 2 & & 34.4332 & $\pm 0.8518$ & 26.7417 & $\pm 0.8388$ \\
                & 3 & & 39.8085 & $\pm 1.5139$ & 32.1170 & $\pm 1.4692$ \\
                & 4 & & 42.9191 & $\pm 1.2776$ & 35.2277 & $\pm 1.2579$ \\
                & 5 & & 39.0510 & $\pm 0.7899$ & 31.3595 & $\pm 0.7147$ \\
                \midrule
                \multirow{5}{10pt}{Frost} & 1 & \multirow{5}{10pt}{7.6914 $\pm 0.1103$} & 13.1572 & $\pm 0.9344$ & 5.4658 & $\pm 1.0086$ \\
                & 2 & & 21.7165 & $\pm 2.2841$ & 14.0250 & $\pm 2.3532$ \\
\bottomrule
            \end{tabular}
        \end{center}
        \label{tab:results_complete_base_corruption_rbft_2}
    \end{minipage}
\end{table}

\begin{table}
    \begin{minipage}{0.49\textwidth}
        \caption{VBFT error rates under individual ImageNet-C data corruption attacks (1-45) with error bars.}
        \tiny
        \begin{center}
            \begin{tabular}{p{0.08\textwidth}p{0.06\textwidth}p{0.1\textwidth}p{0.04\textwidth}p{0.11\textwidth}p{0.04\textwidth}p{0.13\textwidth}}
                \toprule
                Attacks & Severity & $\epsilon_\text{cl}$ & \multicolumn{2}{l}{$\epsilon_\text{at}$} & \multicolumn{2}{l}{$\Delta_\epsilon$} \\
                \midrule
                 \multirow{5}{10pt}{Gaussian Noise} & 1 & \multirow{5}{10pt}{6.9617 $\pm 0.0463$} & 8.0917 & $\pm 0.0993$ & 1.1300 & $\pm 0.0776$ \\
                & 2 & & 9.6622 & $\pm 0.0947$ & 2.7005 & $\pm 0.0657$ \\
                & 3 & & 12.8897 & $\pm 0.3346$ & 5.9281 & $\pm 0.3379$ \\
                & 4 & & 18.1197 & $\pm 0.6454$ & 11.1581 & $\pm 0.6375$ \\
                & 5 & & 25.5728 & $\pm 0.7877$ & 18.6111 & $\pm 0.7758$ \\
                \midrule
                \multirow{5}{10pt}{Shot Noise} & 1 & \multirow{5}{10pt}{6.9617 $\pm 0.0463$} & 7.8070 & $\pm 0.0462$ & 0.8454 & $\pm 0.0295$ \\
                & 2 & & 9.2785 & $\pm 0.0247$ & 2.3168 & $\pm 0.0279$ \\
\bottomrule
            \end{tabular}
        \end{center}
        \label{tab:results_complete_base_corruption_vbft_1}
    \end{minipage}
    \hspace{2pt}
    \begin{minipage}{0.49\textwidth}
        \caption{VBFT error rates under individual ImageNet-C data corruption attacks (46-95) with error bars.}
        \tiny
        \begin{center}
            \begin{tabular}{p{0.08\textwidth}p{0.06\textwidth}p{0.1\textwidth}p{0.04\textwidth}p{0.11\textwidth}p{0.04\textwidth}p{0.13\textwidth}}
                \toprule
                Attacks & Severity & $\epsilon_\text{cl}$ & \multicolumn{2}{l}{$\epsilon_\text{at}$} & \multicolumn{2}{l}{$\Delta_\epsilon$} \\
                \midrule
                \multirow{5}{10pt}{Snow} & 1 & \multirow{5}{10pt}{6.9617 $\pm 0.0463$} & 12.9697 & $\pm 0.9262$ & 6.0080 & $\pm 0.9560$ \\
                & 2 & & 19.0965 & $\pm 0.8283$ & 12.1349 & $\pm 0.8645$ \\
                & 3 & & 24.7043 & $\pm 2.7169$ & 17.7426 & $\pm 2.7582$ \\
                & 4 & & 32.4915 & $\pm 2.9954$ & 25.5298 & $\pm 3.0305$ \\
                & 5 & & 27.7007 & $\pm 1.0346$ & 20.7390 & $\pm 1.0754$ \\
                \midrule
                \multirow{5}{10pt}{Frost} & 1 & \multirow{5}{10pt}{6.9617 $\pm 0.0463$} & 9.9237 & $\pm 0.2336$ & 2.9621 & $\pm 0.2634$ \\
                & 2 & & 15.3228 & $\pm 0.6466$ & 8.3611 & $\pm 0.6337$ \\
\bottomrule
            \end{tabular}
        \end{center}
        \label{tab:results_complete_base_corruption_vbft_2}
    \end{minipage}
\end{table}

\begin{table}
    \caption{MTL model color task error rates under PGD attacks.}
    \scriptsize
    \begin{center}
        \begin{tabular}{lllllllll}
            \toprule
            Models & Attacks & $\epsilon_\text{cl}$ & & $\epsilon_\text{at}$ & & $\Delta_\epsilon$ & \\
            \midrule
            \multirow{8}{*}{RMFT} & RMFT Color & \multirow{8}{*}{8.8558} & \multirow{8}{*}{$\pm 0.1308$} & 97.7737 & $\pm 2.8188$ & 88.9179 & $\pm 2.8236$ \\
            & RMFT Shape & & & 97.7882 & $\pm 2.7967$ & 88.9324 & $\pm 2.8013$ \\
            & RMFT Symbol & & & 97.7730 & $\pm 2.8225$ & 88.9172 & $\pm 2.8265$ \\
            & RMFT Text & & & 97.7651 & $\pm 2.8643$ & 88.9093 & $\pm 2.8687$ \\
            & VMFT Color & & & 97.7737 & $\pm 2.8188$ & 88.9179 & $\pm 2.8236$ \\
            & VMFT Shape & & & 97.7882 & $\pm 2.7967$ & 88.9324 & $\pm 2.8013$ \\
            & VMFT Symbol & & & 97.7730 & $\pm 2.8225$ & 88.9172 & $\pm 2.8265$ \\
\bottomrule
        \end{tabular}
    \end{center}
    \label{tab:results_complete_mtl_pgd_color}
\end{table}

\begin{table}
    \caption{MTL model shape task error rates under PGD attacks.}
    \scriptsize
    \begin{center}
        \begin{tabular}{lllllllll}
            \toprule
            Models & Attacks & $\epsilon_\text{cl}$ & & $\epsilon_\text{at}$ & & $\Delta_\epsilon$ & \\
            \midrule
            \multirow{8}{*}{RMFT} & RMFT Color & \multirow{8}{*}{8.5698} & \multirow{8}{*}{$\pm 0.1548$} & 95.6530 & $\pm 3.8854$ & 87.0832 & $\pm 3.7886$ \\
            & RMFT Shape & & & 95.6603 & $\pm 3.8849$ & 87.0904 & $\pm 3.7893$ \\
            & RMFT Symbol & & & 95.6656 & $\pm 3.8959$ & 87.0957 & $\pm 3.7988$ \\
            & RMFT Text & & & 95.6206 & $\pm 3.9321$ & 87.0508 & $\pm 3.8340$ \\
            & VMFT Color & & & 95.6530 & $\pm 3.8854$ & 87.0832 & $\pm 3.7886$ \\
            & VMFT Shape & & & 95.6603 & $\pm 3.8849$ & 87.0904 & $\pm 3.7893$ \\
            & VMFT Symbol & & & 95.6656 & $\pm 3.8959$ & 87.0957 & $\pm 3.7988$ \\
\bottomrule
        \end{tabular}
    \end{center}
    \label{tab:results_complete_mtl_pgd_shape}
\end{table}

\clearpage

\begin{table}
    \caption{MTL model symbol task error rates under PGD attacks.}
    \scriptsize
    \begin{center}
        \begin{tabular}{lllllllll}
            \toprule
            Models & Attacks & $\epsilon_\text{cl}$ & & $\epsilon_\text{at}$ & & $\Delta_\epsilon$ & \\
            \midrule
            \multirow{8}{*}{RMFT} & RMFT Color & \multirow{8}{*}{11.9645} & \multirow{8}{*}{$\pm 0.3571$} & 99.2603 & $\pm 0.9694$ & 87.2958 & $\pm 0.9963$ \\
            & RMFT Shape & & & 99.2649 & $\pm 0.9613$ & 87.3005 & $\pm 0.9891$ \\
            & RMFT Symbol & & & 99.2649 & $\pm 0.9714$ & 87.3005 & $\pm 1.0017$ \\
            & RMFT Text & & & 99.2557 & $\pm 0.9712$ & 87.2912 & $\pm 1.0022$ \\
            & VMFT Color & & & 99.2603 & $\pm 0.9694$ & 87.2958 & $\pm 0.9963$ \\
            & VMFT Shape & & & 99.2649 & $\pm 0.9613$ & 87.3005 & $\pm 0.9891$ \\
            & VMFT Symbol & & & 99.2649 & $\pm 0.9714$ & 87.3005 & $\pm 1.0017$ \\
\bottomrule
        \end{tabular}
    \end{center}
    \label{tab:results_complete_mtl_pgd_symbol}
\end{table}

\begin{table}
    \caption{MTL model text task error rates under PGD attacks.}
    \scriptsize
    \begin{center}
        \begin{tabular}{lllllllll}
            \toprule
            Models & Attacks & $\epsilon_\text{cl}$ & & $\epsilon_\text{at}$ & & $\Delta_\epsilon$ & \\
            \midrule
            \multirow{8}{*}{RMFT} & RMFT Color & \multirow{8}{*}{4.3912} & \multirow{8}{*}{$\pm 0.1195$} & 99.3561 & $\pm 0.9740$ & 94.9648 & $\pm 1.0668$ \\
            & RMFT Shape & & & 99.3647 & $\pm 0.9640$ & 94.9734 & $\pm 1.0571$ \\
            & RMFT Symbol & & & 99.3508 & $\pm 0.9797$ & 94.9595 & $\pm 1.0726$ \\
            & RMFT Text & & & 99.3614 & $\pm 0.9740$ & 94.9701 & $\pm 1.0665$ \\
            & VMFT Color & & & 99.3561 & $\pm 0.9740$ & 94.9648 & $\pm 1.0668$ \\
            & VMFT Shape & & & 99.3647 & $\pm 0.9640$ & 94.9734 & $\pm 1.0571$ \\
            & VMFT Symbol & & & 99.3508 & $\pm 0.9797$ & 94.9595 & $\pm 1.0726$ \\
\bottomrule
        \end{tabular}
    \end{center}
    \label{tab:results_complete_mtl_pgd_text}
\end{table}

\begin{table}
    \caption{MTL model relative cumulative errors and relative error correlations under PGD attacks.}
    \scriptsize
    \begin{center}
        \begin{tabular}{llllll}
            \toprule
            Models & Attacks & \multicolumn{2}{l}{$\epsilon_\text{cu}$} & \multicolumn{2}{l}{\textit{RECorr}} \\
            \midrule
            \multirow{8}{*}{RMFT} & RMFT Color & 358.2617 & $\pm 3.6374$ & 1.0110 & $\pm 0.0423$ \\
            & RMFT Shape & 358.2967 & $\pm 3.6004$ & 1.0404 & $\pm 0.0547$ \\
            & RMFT Symbol & 358.2730 & $\pm 3.6471$ & 1.0347 & $\pm 0.0157$ \\
            & RMFT Text & 358.2214 & $\pm 3.6738$ & 0.9241 & $\pm 0.0155$ \\
            & VMFT Color & 358.1686 & $\pm 3.6538$ & 1.0161 & $\pm 0.0451$ \\
            & VMFT Shape & 358.1468 & $\pm 3.7001$ & 1.0392 & $\pm 0.0521$ \\
            & VMFT Symbol & 358.1244 & $\pm 3.6557$ & 1.0320 & $\pm 0.0128$ \\
\bottomrule
        \end{tabular}
    \end{center}
    \label{tab:results_complete_mtl_pgd_cumulative}
\end{table}

\begin{table}
    \begin{minipage}{0.49\textwidth}
        \caption{RMFT color task error rates under individual ImageNet-C data corruption attacks (1-45) with error bars.}
        \tiny
        \begin{center}
            \begin{tabular}
                {p{0.08\textwidth}p{0.06\textwidth}p{0.1\textwidth}p{0.04\textwidth}p{0.11\textwidth}p{0.04\textwidth}p{0.13\textwidth}}
                \toprule
                Attacks & Severity & $\epsilon_\text{cl}$ & \multicolumn{2}{l}{$\epsilon_\text{at}$} & \multicolumn{2}{l}{$\Delta_\epsilon$} \\
                \midrule
                \multirow{5}{10pt}{Gaussian Noise} & 1 & \multirow{5}{10pt}{8.8558 $\pm 0.1308$} & 22.1794 & $\pm 2.3228$ & 13.3236 & $\pm 2.4133$ \\
                & 2 & & 37.3107 & $\pm 3.9399$ & 28.4549 & $\pm 4.0185$ \\
                & 3 & & 47.5944 & $\pm 1.7283$ & 38.7386 & $\pm 1.7913$ \\
                & 4 & & 50.7156 & $\pm 0.3472$ & 41.8598 & $\pm 0.3971$ \\
                & 5 & & 51.2274 & $\pm 0.0255$ & 42.3716 & $\pm 0.1352$ \\
                \midrule
                \multirow{5}{10pt}{Shot Noise} & 1 & \multirow{5}{10pt}{8.8558 $\pm 0.1308$} & 22.5539 & $\pm 1.6901$ & 13.6981 & $\pm 1.7821$ \\
                & 2 & & 36.9666 & $\pm 3.2683$ & 28.1108 & $\pm 3.3446$ \\
\bottomrule
            \end{tabular}
        \end{center}
        \label{tab:results_complete_mtl_corruption_rmft_color_1}
    \end{minipage}
    \hspace{2pt}
    \begin{minipage}{0.49\textwidth}
        \caption{RMFT color task error rates under individual ImageNet-C data corruption attacks (46-95) with error bars.}
        \tiny
        \begin{center}
            \begin{tabular}{p{0.08\textwidth}p{0.06\textwidth}p{0.1\textwidth}p{0.04\textwidth}p{0.11\textwidth}p{0.04\textwidth}p{0.13\textwidth}}
                \toprule
                Attacks & Severity & $\epsilon_\text{cl}$ & \multicolumn{2}{l}{$\epsilon_\text{at}$} & \multicolumn{2}{l}{$\Delta_\epsilon$} \\
                \midrule
                \multirow{5}{10pt}{Snow} & 1 & \multirow{5}{10pt}{8.8558 $\pm 0.1308$} & 35.9674 & $\pm 1.2763$ & 27.1116 & $\pm 1.2033$ \\
                & 2 & & 41.3341 & $\pm 1.1933$ & 32.4783 & $\pm 1.1160$ \\
                & 3 & & 44.5814 & $\pm 0.7821$ & 35.7257 & $\pm 0.7135$ \\
                & 4 & & 46.9544 & $\pm 0.5323$ & 38.0986 & $\pm 0.4489$ \\
                & 5 & & 45.5536 & $\pm 1.2083$ & 36.6978 & $\pm 1.1050$ \\
                \midrule
                \multirow{5}{10pt}{Frost} & 1 & \multirow{5}{10pt}{8.8558 $\pm 0.1308$} & 17.7737 & $\pm 0.6782$ & 8.9179 & $\pm 0.6003$ \\
                & 2 & & 28.0243 & $\pm 0.8164$ & 19.1685 & $\pm 0.7493$ \\
\bottomrule
            \end{tabular}
        \end{center}
        \label{tab:results_complete_mtl_corruption_rmft_color_2}
    \end{minipage}
\end{table}

\begin{table}
    \begin{minipage}{0.49\textwidth}
        \caption{RMFT shape task error rates under individual ImageNet-C data corruption attacks (1-45) with error bars.}
        \tiny
        \begin{center}
            \begin{tabular}{p{0.08\textwidth}p{0.06\textwidth}p{0.1\textwidth}p{0.04\textwidth}p{0.11\textwidth}p{0.04\textwidth}p{0.13\textwidth}}
                \toprule
                Attacks & Severity & $\epsilon_\text{cl}$ & \multicolumn{2}{l}{$\epsilon_\text{at}$} & \multicolumn{2}{l}{$\Delta_\epsilon$} \\
                \midrule
                \multirow{5}{10pt}{Gaussian Noise} & 1 & \multirow{5}{10pt}{8.5698 $\pm 0.1548$} & 21.7257 & $\pm 2.3047$ & 13.1559 & $\pm 2.4293$ \\
                & 2 & & 36.3029 & $\pm 3.7999$ & 27.7331 & $\pm 3.9090$ \\
                & 3 & & 46.8646 & $\pm 1.9772$ & 38.2948 & $\pm 2.0713$ \\
                & 4 & & 50.5221 & $\pm 0.4733$ & 41.9523 & $\pm 0.5434$ \\
                & 5 & & 51.2202 & $\pm 0.0413$ & 42.6503 & $\pm 0.1695$ \\
                \midrule
                \multirow{5}{10pt}{Shot Noise} & 1 & \multirow{5}{10pt}{8.5698 $\pm 0.1548$} & 22.1286 & $\pm 1.5253$ & 13.5588 & $\pm 1.6251$ \\
                & 2 & & 36.1067 & $\pm 2.9796$ & 27.5369 & $\pm 3.0821$ \\
\bottomrule
            \end{tabular}
        \end{center}
        \label{tab:results_complete_mtl_corruption_rmft_shape_1}
    \end{minipage}
    \hspace{2pt}
    \begin{minipage}{0.49\textwidth}
        \caption{RMFT shape task error rates under individual ImageNet-C data corruption attacks (46-95) with error bars.}
        \tiny
        \begin{center}
            \begin{tabular}{p{0.08\textwidth}p{0.06\textwidth}p{0.1\textwidth}p{0.04\textwidth}p{0.11\textwidth}p{0.04\textwidth}p{0.13\textwidth}}
                \toprule
                Attacks & Severity & $\epsilon_\text{cl}$ & \multicolumn{2}{l}{$\epsilon_\text{at}$} & \multicolumn{2}{l}{$\Delta_\epsilon$} \\
                \midrule
                \multirow{5}{10pt}{Snow} & 1 & \multirow{5}{10pt}{8.5698 $\pm 0.1548$} & 35.5341 & $\pm 1.3769$ & 26.9643 & $\pm 1.2551$ \\
                & 2 & & 40.9517 & $\pm 1.4060$ & 32.3819 & $\pm 1.2821$ \\
                & 3 & & 44.5009 & $\pm 0.9047$ & 35.9311 & $\pm 0.7863$ \\
                & 4 & & 46.9108 & $\pm 0.6432$ & 38.3410 & $\pm 0.5240$ \\
                & 5 & & 45.3918 & $\pm 1.3557$ & 36.8220 & $\pm 1.2158$ \\
                \midrule
                \multirow{5}{10pt}{Frost} & 1 & \multirow{5}{10pt}{8.5698 $\pm 0.1548$} & 17.7175 & $\pm 0.7676$ & 9.1477 & $\pm 0.6419$ \\
                & 2 & & 27.7912 & $\pm 0.8774$ & 19.2213 & $\pm 0.7516$ \\
\bottomrule
            \end{tabular}
        \end{center}
        \label{tab:results_complete_mtl_corruption_rmft_shape_2}
    \end{minipage}
\end{table}

\begin{table}
    \begin{minipage}{0.49\textwidth}
        \caption{RMFT symbol task error rates under individual ImageNet-C data corruption attacks (1-45) with error bars.}
        \tiny
        \begin{center}
            \begin{tabular}{p{0.08\textwidth}p{0.06\textwidth}p{0.1\textwidth}p{0.04\textwidth}p{0.11\textwidth}p{0.04\textwidth}p{0.13\textwidth}}
                \toprule
                Attacks & Severity & $\epsilon_\text{cl}$ & \multicolumn{2}{l}{$\epsilon_\text{at}$} & \multicolumn{2}{l}{$\Delta_\epsilon$} \\
                \midrule
                \multirow{5}{10pt}{Gaussian Noise} & 1 & \multirow{5}{10pt}{11.9645 $\pm 0.3571$} & 23.7354 & $\pm 1.3828$ & 11.7710 & $\pm 1.7135$ \\
                & 2 & & 32.8726 & $\pm 1.8745$ & 20.9081 & $\pm 2.1971$ \\
                & 3 & & 38.0200 & $\pm 0.6990$ & 26.0555 & $\pm 1.0037$ \\
                & 4 & & 39.1216 & $\pm 0.1017$ & 27.1572 & $\pm 0.4295$ \\
                & 5 & & 39.2313 & $\pm 0.0040$ & 27.2668 & $\pm 0.3589$ \\
                \midrule
                \multirow{5}{10pt}{Shot Noise} & 1 & \multirow{5}{10pt}{11.9645 $\pm 0.3571$} & 24.1627 & $\pm 0.8860$ & 12.1983 & $\pm 1.1594$ \\
                & 2 & & 32.8019 & $\pm 1.5330$ & 20.8374 & $\pm 1.8373$ \\
\bottomrule
            \end{tabular}
        \end{center}
        \label{tab:results_complete_mtl_corruption_rmft_symbol_1}
    \end{minipage}
    \hspace{2pt}
    \begin{minipage}{0.49\textwidth}
        \caption{RMFT symbol task error rates under individual ImageNet-C data corruption attacks (46-95) with error bars.}
        \tiny
        \begin{center}
            \begin{tabular}{p{0.08\textwidth}p{0.06\textwidth}p{0.1\textwidth}p{0.04\textwidth}p{0.11\textwidth}p{0.04\textwidth}p{0.13\textwidth}}
                \toprule
                Attacks & Severity & $\epsilon_\text{cl}$ & \multicolumn{2}{l}{$\epsilon_\text{at}$} & \multicolumn{2}{l}{$\Delta_\epsilon$} \\
                \midrule
                \multirow{5}{10pt}{Snow} & 1 & \multirow{5}{10pt}{11.9645 $\pm 0.3571$} & 32.6196 & $\pm 0.6838$ & 20.6552 & $\pm 0.5017$ \\
                & 2 & & 35.6187 & $\pm 0.5333$ & 23.6542 & $\pm 0.4911$ \\
                & 3 & & 36.9012 & $\pm 0.3387$ & 24.9368 & $\pm 0.3177$ \\
                & 4 & & 38.1072 & $\pm 0.1717$ & 26.1427 & $\pm 0.2757$ \\
                & 5 & & 37.3153 & $\pm 0.4424$ & 25.3509 & $\pm 0.2779$ \\
                \midrule
                \multirow{5}{10pt}{Frost} & 1 & \multirow{5}{10pt}{11.9645 $\pm 0.3571$} & 20.7575 & $\pm 0.7359$ & 8.7931 & $\pm 0.3933$ \\
                & 2 & & 28.3241 & $\pm 0.8716$ & 16.3597 & $\pm 0.5530$ \\
\bottomrule
            \end{tabular}
        \end{center}
        \label{tab:results_complete_mtl_corruption_rmft_symbol_2}
    \end{minipage}
\end{table}

\begin{table}
    \begin{minipage}{0.49\textwidth}
        \caption{RMFT text task error rates under individual ImageNet-C data corruption attacks (1-45) with error bars.}
        \tiny
        \begin{center}
            \begin{tabular}{p{0.08\textwidth}p{0.06\textwidth}p{0.1\textwidth}p{0.04\textwidth}p{0.11\textwidth}p{0.04\textwidth}p{0.13\textwidth}}
                \toprule
                Attacks & Severity & $\epsilon_\text{cl}$ & \multicolumn{2}{l}{$\epsilon_\text{at}$} & \multicolumn{2}{l}{$\Delta_\epsilon$} \\
                \midrule
                \multirow{5}{10pt}{Gaussian Noise} & 1 & \multirow{5}{10pt}{4.3912 $\pm 0.1195$} & 9.5215 & $\pm 0.2916$ & 5.1303 & $\pm 0.4013$ \\
                & 2 & & 12.9076 & $\pm 0.5924$ & 8.5163 & $\pm 0.7038$ \\
                & 3 & & 14.9728 & $\pm 0.3069$ & 10.5815 & $\pm 0.4153$ \\
                & 4 & & 15.5117 & $\pm 0.0471$ & 11.1204 & $\pm 0.1626$ \\
                & 5 & & 15.5658 & $\pm 0.0013$ & 11.1746 & $\pm 0.1205$ \\
                \midrule
                \multirow{5}{10pt}{Shot Noise} & 1 & \multirow{5}{10pt}{4.3912 $\pm 0.1195$} & 10.5241 & $\pm 0.3647$ & 6.1328 & $\pm 0.4063$ \\
                & 2 & & 13.3600 & $\pm 0.4505$ & 8.9687 & $\pm 0.5378$ \\
\bottomrule
            \end{tabular}
        \end{center}
        \label{tab:results_complete_mtl_corruption_rmft_text_1}
    \end{minipage}
    \hspace{2pt}
    \begin{minipage}{0.49\textwidth}
        \caption{RMFT text task error rates under individual ImageNet-C data corruption attacks (46-95) with error bars.}
        \tiny
        \begin{center}
            \begin{tabular}{p{0.08\textwidth}p{0.06\textwidth}p{0.1\textwidth}p{0.04\textwidth}p{0.11\textwidth}p{0.04\textwidth}p{0.13\textwidth}}
                \toprule
                Attacks & Severity & $\epsilon_\text{cl}$ & \multicolumn{2}{l}{$\epsilon_\text{at}$} & \multicolumn{2}{l}{$\Delta_\epsilon$} \\
                \midrule
                \multirow{5}{10pt}{Snow} & 1 & \multirow{5}{10pt}{4.3912 $\pm 0.1195$} & 12.6454 & $\pm 0.3667$ & 8.2541 & $\pm 0.3126$ \\
                & 2 & & 13.8441 & $\pm 0.3157$ & 9.4528 & $\pm 0.2970$ \\
                & 3 & & 14.4213 & $\pm 0.2163$ & 10.0300 & $\pm 0.1658$ \\
                & 4 & & 14.9278 & $\pm 0.1341$ & 10.5366 & $\pm 0.1197$ \\
                & 5 & & 14.5857 & $\pm 0.3082$ & 10.1945 & $\pm 0.2553$ \\
                \midrule
                \multirow{5}{10pt}{Frost} & 1 & \multirow{5}{10pt}{4.3912 $\pm 0.1195$} & 8.5203 & $\pm 0.3097$ & 4.1290 & $\pm 0.2412$ \\
                & 2 & & 11.7822 & $\pm 0.3876$ & 7.3909 & $\pm 0.3350$ \\
\bottomrule
            \end{tabular}
        \end{center}
        \label{tab:results_complete_mtl_corruption_rmft_text_2}
    \end{minipage}
\end{table}

\begin{table}
    \begin{minipage}{0.48\textwidth}
        \caption{RMFT cumulative relative errors under individual ImageNet-C data corruption attacks (1-45) with error bars.}
        \scriptsize
        \begin{center}
            \begin{tabular}{llll}
                \toprule
                Attacks & Severity & \multicolumn{2}{l}{$\epsilon_\text{cu}$} \\
                \midrule
                \multirow{5}{10pt}{Gaussian Noise} & 1 & 43.3808 & $\pm 6.9290$ \\
                & 2 & 85.6124 & $\pm 10.7916$ \\
                & 3 & 113.6704 & $\pm 5.2531$ \\
                & 4 & 122.0896 & $\pm 1.4926$ \\
                & 5 & 123.4633 & $\pm 0.7564$ \\
                \midrule
                \multirow{5}{10pt}{Shot Noise} & 1 & 45.5880 & $\pm 4.8971$ \\
                & 2 & 85.4539 & $\pm 8.7284$ \\
\bottomrule
            \end{tabular}
        \end{center}
        \label{tab:results_complete_mtl_corruption_rmft_cumulative_1}
    \end{minipage}
    \quad
    \begin{minipage}{0.48\textwidth}
        \caption{RMFT cumulative relative errors under individual ImageNet-C data corruption attacks (46-95) with error bars.}
        \scriptsize
        \begin{center}
            \begin{tabular}{llll}
                \toprule
                Attacks & Severity & \multicolumn{2}{l}{$\epsilon_\text{cu}$} \\
                \midrule
                \multirow{5}{10pt}{Snow} & 1 & 82.9852 & $\pm 3.2059$ \\
                & 2 & 97.9672 & $\pm 2.9755$ \\
                & 3 & 106.6235 & $\pm 1.7593$ \\
                & 4 & 113.1189 & $\pm 0.9614$ \\
                & 5 & 109.0652 & $\pm 2.6937$ \\
                \midrule
                \multirow{5}{10pt}{Frost} & 1 & 30.9877 & $\pm 1.8200$ \\
                & 2 & 62.1405 & $\pm 2.2907$ \\
\bottomrule
            \end{tabular}
        \end{center}
        \label{tab:results_complete_mtl_corruption_rmft_cumulative_2}
    \end{minipage}
\end{table}

\begin{table}
    \begin{minipage}{0.49\textwidth}
        \caption{VMFT color task error rates under individual ImageNet-C data corruption attacks (1-45) with error bars.}
    \tiny
    \begin{center}
        \begin{tabular}{p{0.08\textwidth}p{0.06\textwidth}p{0.1\textwidth}p{0.04\textwidth}p{0.11\textwidth}p{0.04\textwidth}p{0.13\textwidth}}
            \toprule
            Attacks & Severity & $\epsilon_\text{cl}$ & \multicolumn{2}{l}{$\epsilon_\text{at}$} & \multicolumn{2}{l}{$\Delta_\epsilon$} \\
            \midrule
            \multirow{5}{10pt}{Gaussian Noise} & 1 & \multirow{5}{10pt}{7.6914 $\pm 0.1041$} & 10.1958 & $\pm 0.2431$ & 2.5044 & $\pm 0.2945$ \\
            & 2 & & 13.0370 & $\pm 0.4871$ & 5.3456 & $\pm 0.5686$ \\
            & 3 & & 17.5656 & $\pm 0.9523$ & 9.8742 & $\pm 1.0222$ \\
            & 4 & & 23.3458 & $\pm 1.4776$ & 15.6543 & $\pm 1.5336$ \\
            & 5 & & 31.0914 & $\pm 1.7558$ & 23.3999 & $\pm 1.8150$ \\
            \midrule
            \multirow{5}{10pt}{Shot Noise} & 1 & \multirow{5}{10pt}{7.6914 $\pm 0.1041$} & 9.9198 & $\pm 0.2451$ & 2.2283 & $\pm 0.3414$ \\
            & 2 & & 12.7933 & $\pm 0.5463$ & 5.1019 & $\pm 0.6257$ \\
\bottomrule
            \end{tabular}
        \end{center}
        \label{tab:results_complete_mtl_corruption_vmft_color_1}
    \end{minipage}
    \hspace{2pt}
    \begin{minipage}{0.49\textwidth}
         \caption{VMFT color task error rates under individual ImageNet-C data corruption attacks (46-95) with error bars.}
        \tiny
        \begin{center}
            \begin{tabular}{p{0.08\textwidth}p{0.06\textwidth}p{0.1\textwidth}p{0.04\textwidth}p{0.11\textwidth}p{0.04\textwidth}p{0.13\textwidth}}
                \toprule
                Attacks & Severity & $\epsilon_\text{cl}$ & \multicolumn{2}{l}{$\epsilon_\text{at}$} & \multicolumn{2}{l}{$\Delta_\epsilon$} \\
                \midrule
                \multirow{5}{10pt}{Snow} & 1 & \multirow{5}{10pt}{7.6914 $\pm 0.1041$} & 17.6957 & $\pm 0.6952$ & 10.0043 & $\pm 0.6563$ \\
                & 2 & & 25.0913 & $\pm 0.8483$ & 17.3999 & $\pm 0.9003$ \\
                & 3 & & 28.4893 & $\pm 0.9757$ & 20.7978 & $\pm 0.9646$ \\
                & 4 & & 36.4581 & $\pm 1.1642$ & 28.7666 & $\pm 1.2099$ \\
                & 5 & & 34.0772 & $\pm 1.1558$ & 26.3858 & $\pm 1.1776$ \\
                \midrule
                \multirow{5}{10pt}{Frost} & 1 & \multirow{5}{10pt}{7.6914 $\pm 0.1041$} & 11.1832 & $\pm 0.1355$ & 3.4917 & $\pm 0.1691$ \\
                & 2 & & 17.9969 & $\pm 0.2447$ & 10.3055 & $\pm 0.3219$ \\
\bottomrule
            \end{tabular}
        \end{center}
        \label{tab:results_complete_mtl_corruption_vmft_color_2}
    \end{minipage}
\end{table}

\begin{table}
    \begin{minipage}{0.49\textwidth}
        \caption{VMFT shape task error rates under individual ImageNet-C data corruption attacks (1-45) with error bars.}
        \tiny
        \begin{center}
            \begin{tabular}{p{0.08\textwidth}p{0.06\textwidth}p{0.1\textwidth}p{0.04\textwidth}p{0.11\textwidth}p{0.04\textwidth}p{0.13\textwidth}}
                \toprule
                Attacks & Severity & $\epsilon_\text{cl}$ & \multicolumn{2}{l}{$\epsilon_\text{at}$} & \multicolumn{2}{l}{$\Delta_\epsilon$} \\
                \midrule
                \multirow{5}{10pt}{Gaussian Noise} & 1 & \multirow{5}{10pt}{7.5950 $\pm 0.0763$} & 9.9171 & $\pm 0.2765$ & 2.3221 & $\pm 0.2796$ \\
                & 2 & & 12.5186 & $\pm 0.5772$ & 4.9236 & $\pm 0.6134$ \\
                & 3 & & 16.8306 & $\pm 1.0520$ & 9.2355 & $\pm 1.0971$ \\
                & 4 & & 22.4264 & $\pm 1.6316$ & 14.8314 & $\pm 1.6739$ \\
                & 5 & & 30.1073 & $\pm 1.8577$ & 22.5123 & $\pm 1.9019$ \\
                \midrule
                \multirow{5}{10pt}{Shot Noise} & 1 & \multirow{5}{10pt}{7.5950 $\pm 0.0763$} & 9.6820 & $\pm 0.2586$ & 2.0870 & $\pm 0.2886$ \\
                & 2 & & 12.3158 & $\pm 0.6126$ & 4.7208 & $\pm 0.6517$ \\
\bottomrule
            \end{tabular}
        \end{center}
        \label{tab:results_complete_mtl_corruption_vmft_shape_1}
    \end{minipage}
    \hspace{2pt}
    \begin{minipage}{0.49\textwidth}
        \caption{VMFT shape task error rates under individual ImageNet-C data corruption attacks (46-95) with error bars.}
        \tiny
        \begin{center}
            \begin{tabular}{p{0.08\textwidth}p{0.06\textwidth}p{0.1\textwidth}p{0.04\textwidth}p{0.11\textwidth}p{0.04\textwidth}p{0.13\textwidth}}
                \toprule
                Attacks & Severity & $\epsilon_\text{cl}$ & \multicolumn{2}{l}{$\epsilon_\text{at}$} & \multicolumn{2}{l}{$\Delta_\epsilon$} \\
                \midrule
                \multirow{5}{10pt}{Snow} & 1 & \multirow{5}{10pt}{7.5950 $\pm 0.0763$} & 17.2031 & $\pm 0.8645$ & 9.6080 & $\pm 0.8249$ \\
                & 2 & & 24.4520 & $\pm 0.9966$ & 16.8570 & $\pm 0.9817$ \\
                & 3 & & 28.0197 & $\pm 1.1027$ & 20.4247 & $\pm 1.0782$ \\
                & 4 & & 36.0341 & $\pm 1.2586$ & 28.4391 & $\pm 1.2559$ \\
                & 5 & & 33.3408 & $\pm 1.3892$ & 25.7458 & $\pm 1.3811$ \\
                \midrule
                \multirow{5}{10pt}{Frost} & 1 & \multirow{5}{10pt}{7.5950 $\pm 0.0763$} & 10.7981 & $\pm 0.1878$ & 3.2031 & $\pm 0.1595$ \\
                & 2 & & 17.4263 & $\pm 0.2168$ & 9.8313 & $\pm 0.2201$ \\
\bottomrule
            \end{tabular}
        \end{center}
        \label{tab:results_complete_mtl_corruption_vmft_shape_2}
    \end{minipage}
\end{table}

\begin{table}
    \begin{minipage}{0.49\textwidth}
        \caption{VMFT symbol task error rates under individual ImageNet-C data corruption attacks (1-45) with error bars.}
        \tiny
        \begin{center}
            \begin{tabular}{p{0.08\textwidth}p{0.06\textwidth}p{0.1\textwidth}p{0.04\textwidth}p{0.11\textwidth}p{0.04\textwidth}p{0.13\textwidth}}
                \toprule
                Attacks & Severity & $\epsilon_\text{cl}$ & \multicolumn{2}{l}{$\epsilon_\text{at}$} & \multicolumn{2}{l}{$\Delta_\epsilon$} \\
                \midrule
                \multirow{5}{10pt}{Gaussian Noise} & 1 & \multirow{5}{10pt}{11.0768 $\pm 0.2036$} & 13.4267 & $\pm 0.2380$ & 2.3498 & $\pm 0.1912$ \\
                & 2 & & 15.7725 & $\pm 0.5478$ & 4.6957 & $\pm 0.4619$ \\
                & 3 & & 19.2379 & $\pm 0.9106$ & 8.1610 & $\pm 0.8087$ \\
                & 4 & & 23.4158 & $\pm 1.2740$ & 12.3389 & $\pm 1.1908$ \\
                & 5 & & 28.6332 & $\pm 1.2642$ & 17.5564 & $\pm 1.1852$ \\
                \midrule
                \multirow{5}{10pt}{Shot Noise} & 1 & \multirow{5}{10pt}{11.0768 $\pm 0.2036$} & 13.1850 & $\pm 0.3646$ & 2.1081 & $\pm 0.2684$ \\
                & 2 & & 15.5354 & $\pm 0.5449$ & 4.4586 & $\pm 0.4815$ \\
\bottomrule
            \end{tabular}
        \end{center}
        \label{tab:results_complete_mtl_corruption_vmft_symbol_1}
    \end{minipage}
    \hspace{2pt}
    \begin{minipage}{0.49\textwidth}
        \caption{VMFT symbol task error rates under individual ImageNet-C data corruption attacks (46-95) with error bars.}
        \tiny
        \begin{center}
            \begin{tabular}{p{0.08\textwidth}p{0.06\textwidth}p{0.1\textwidth}p{0.04\textwidth}p{0.11\textwidth}p{0.04\textwidth}p{0.13\textwidth}}
                \toprule
                Attacks & Severity & $\epsilon_\text{cl}$ & \multicolumn{2}{l}{$\epsilon_\text{at}$} & \multicolumn{2}{l}{$\Delta_\epsilon$} \\
                \midrule
                \multirow{5}{10pt}{Snow} & 1 & \multirow{5}{10pt}{11.0768 $\pm 0.2036$} & 19.7299 & $\pm 0.6700$ & 8.6530 & $\pm 0.5250$ \\
                & 2 & & 24.8621 & $\pm 0.4472$ & 13.7853 & $\pm 0.3897$ \\
                & 3 & & 27.1651 & $\pm 0.6893$ & 16.0882 & $\pm 0.7143$ \\
                & 4 & & 32.4492 & $\pm 0.5427$ & 21.3724 & $\pm 0.6713$ \\
                & 5 & & 31.1587 & $\pm 2.1779$ & 20.0819 & $\pm 2.1720$ \\
                \midrule
                \multirow{5}{10pt}{Frost} & 1 & \multirow{5}{10pt}{11.0768 $\pm 0.2036$} & 14.3586 & $\pm 0.2197$ & 3.2817 & $\pm 0.1949$ \\
                & 2 & & 19.7081 & $\pm 0.3416$ & 8.6312 & $\pm 0.3123$ \\
\bottomrule
            \end{tabular}
        \end{center}
        \label{tab:results_complete_mtl_corruption_vmft_symbol_2}
    \end{minipage}
\end{table}

\begin{table}
    \begin{minipage}{0.49\textwidth}
        \caption{VMFT text task error rates under individual ImageNet-C data corruption attacks (1-45) with error bars.}
        \tiny
        \begin{center}
            \begin{tabular}{p{0.08\textwidth}p{0.06\textwidth}p{0.1\textwidth}p{0.04\textwidth}p{0.11\textwidth}p{0.04\textwidth}p{0.13\textwidth}}
                \toprule
                Attacks & Severity & $\epsilon_\text{cl}$ & \multicolumn{2}{l}{$\epsilon_\text{at}$} & \multicolumn{2}{l}{$\Delta_\epsilon$} \\
                \midrule
                \multirow{5}{10pt}{Gaussian Noise} & 1 & \multirow{5}{10pt}{3.9970 $\pm 0.1423$} & 5.1422 & $\pm 0.1444$ & 1.1452 & $\pm 0.0519$ \\
                & 2 & & 6.3501 & $\pm 0.2531$ & 2.3531 & $\pm 0.1485$ \\
                & 3 & & 8.0329 & $\pm 0.3269$ & 4.0359 & $\pm 0.2471$ \\
                & 4 & & 9.9898 & $\pm 0.5239$ & 5.9928 & $\pm 0.4592$ \\
                & 5 & & 12.1098 & $\pm 0.4893$ & 8.1128 & $\pm 0.4641$ \\
                \midrule
                \multirow{5}{10pt}{Shot Noise} & 1 & \multirow{5}{10pt}{3.9970 $\pm 0.1423$} & 5.1706 & $\pm 0.1640$ & 1.1736 & $\pm 0.0606$ \\
                & 2 & & 6.4875 & $\pm 0.2349$ & 2.4905 & $\pm 0.1494$ \\
\bottomrule
            \end{tabular}
        \end{center}
        \label{tab:results_complete_mtl_corruption_vmft_text_1}
    \end{minipage}
    \hspace{2pt}
    \begin{minipage}{0.49\textwidth}
        \caption{VMFT text task error rates under individual ImageNet-C data corruption attacks (46-95) with error bars.}
        \tiny
        \begin{center}
            \begin{tabular}{p{0.08\textwidth}p{0.06\textwidth}p{0.1\textwidth}p{0.04\textwidth}p{0.11\textwidth}p{0.04\textwidth}p{0.13\textwidth}}
                \toprule
                Attacks & Severity & $\epsilon_\text{cl}$ & \multicolumn{2}{l}{$\epsilon_\text{at}$} & \multicolumn{2}{l}{$\Delta_\epsilon$} \\
                \midrule
                \multirow{5}{10pt}{Snow} & 1 & \multirow{5}{10pt}{3.9970 $\pm 0.1423$} & 7.7826 & $\pm 0.3320$ & 3.7856 & $\pm 0.3056$ \\
                & 2 & & 10.2335 & $\pm 0.4873$ & 6.2365 & $\pm 0.5108$ \\
                & 3 & & 10.7876 & $\pm 0.3019$ & 6.7906 & $\pm 0.2577$ \\
                & 4 & & 12.9723 & $\pm 0.2789$ & 8.9753 & $\pm 0.3105$ \\
                & 5 & & 11.9116 & $\pm 0.3396$ & 7.9147 & $\pm 0.3279$ \\
                \midrule
                \multirow{5}{10pt}{Frost} & 1 & \multirow{5}{10pt}{3.9970 $\pm 0.1423$} & 5.4261 & $\pm 0.1265$ & 1.4292 & $\pm 0.0622$ \\
                & 2 & & 7.6789 & $\pm 0.2472$ & 3.6819 & $\pm 0.1869$ \\
\bottomrule
            \end{tabular}
        \end{center}
        \label{tab:results_complete_mtl_corruption_vmft_text_2}
    \end{minipage}
\end{table}

\begin{table}
    \begin{minipage}{0.48\textwidth}
        \caption{VMFT cumulative relative errors under individual ImageNet-C data corruption attacks (1-45) with error bars.}
        \scriptsize
        \begin{center}
            \begin{tabular}{llll}
                \toprule
                Attacks & Severity & \multicolumn{2}{l}{$\epsilon_\text{cu}$} \\
                \midrule
                \multirow{5}{10pt}{Gaussian Noise} & 1 & 8.3215 & $\pm 0.7621$ \\
                & 2 & 17.3180 & $\pm 1.6904$ \\
                & 3 & 31.3067 & $\pm 3.0588$ \\
                & 4 & 48.8175 & $\pm 4.7506$ \\
                & 5 & 71.5814 & $\pm 5.2377$ \\
                \midrule
                \multirow{5}{10pt}{Shot Noise} & 1 & 7.5970 & $\pm 0.8612$ \\
                & 2 & 16.7718 & $\pm 1.8549$ \\
\bottomrule
            \end{tabular}
        \end{center}
        \label{tab:results_complete_mtl_corruption_vmft_cumulative_1}
    \end{minipage}
    \quad
    \begin{minipage}{0.48\textwidth}
        \caption{VMFT cumulative relative errors under individual ImageNet-C data corruption attacks (46-95) with error bars.}
        \scriptsize
        \begin{center}
            \begin{tabular}{llll}
                \toprule
                Attacks & Severity & \multicolumn{2}{l}{$\epsilon_\text{cu}$} \\
                \midrule
                \multirow{5}{10pt}{Snow} & 1 & 32.0510 & $\pm 2.1843$ \\
                & 2 & 54.2786 & $\pm 2.6809$ \\
                & 3 & 64.1013 & $\pm 2.9287$ \\
                & 4 & 87.5534 & $\pm 3.4018$ \\
                & 5 & 80.1281 & $\pm 4.8348$ \\
                \midrule
                \multirow{5}{10pt}{Frost} & 1 & 11.4057 & $\pm 0.4556$ \\
                & 2 & 32.4499 & $\pm 0.7693$ \\
\bottomrule
            \end{tabular}
        \end{center}
        \label{tab:results_complete_mtl_corruption_vmft_cumulative_2}
    \end{minipage}
\end{table}

\begin{table}
    \begin{minipage}{0.49\textwidth}
        \caption{RMFTL color task error rates under individual ImageNet-C data corruption attacks (1-45) with error bars.}
        \tiny
        \begin{center}
            \begin{tabular}{p{0.08\textwidth}p{0.06\textwidth}p{0.1\textwidth}p{0.04\textwidth}p{0.11\textwidth}p{0.04\textwidth}p{0.13\textwidth}}
                \toprule
                Attacks & Severity & $\epsilon_\text{cl}$ & \multicolumn{2}{l}{$\epsilon_\text{at}$} & \multicolumn{2}{l}{$\Delta_\epsilon$} \\
                \midrule
                \multirow{5}{10pt}{Gaussian Noise} & 1 & \multirow{5}{10pt}{27.8757 $\pm 0.6371$} & 51.2400 & $\pm 0.0075$ & 23.3643 & $\pm 0.6390$ \\
                & 2 & & 51.2466 & $\pm 0.0000$ & 23.3709 & $\pm 0.6371$ \\
                & 3 & & 51.2466 & $\pm 0.0000$ & 23.3709 & $\pm 0.6371$ \\
                & 4 & & 51.2466 & $\pm 0.0000$ & 23.3709 & $\pm 0.6371$ \\
                & 5 & & 51.2466 & $\pm 0.0000$ & 23.3709 & $\pm 0.6371$ \\
                \midrule
                \multirow{5}{10pt}{Shot Noise} & 1 & \multirow{5}{10pt}{27.8757 $\pm 0.6371$} & 51.1614 & $\pm 0.0247$ & 23.2857 & $\pm 0.6307$ \\
                & 2 & & 51.2466 & $\pm 0.0000$ & 23.3709 & $\pm 0.6371$ \\
\bottomrule
            \end{tabular}
        \end{center}
        \label{tab:results_complete_mtl_corruption_rmftl_color_1}
    \end{minipage}
    \hspace{2pt}
    \begin{minipage}{0.49\textwidth}
        \caption{RMFTL color task error rates under individual ImageNet-C data corruption attacks (46-95) with error bars.}
        \tiny
        \begin{center}
            \begin{tabular}{p{0.08\textwidth}p{0.06\textwidth}p{0.1\textwidth}p{0.04\textwidth}p{0.11\textwidth}p{0.04\textwidth}p{0.13\textwidth}}
                \toprule
                Attacks & Severity & $\epsilon_\text{cl}$ & \multicolumn{2}{l}{$\epsilon_\text{at}$} & \multicolumn{2}{l}{$\Delta_\epsilon$} \\
                \midrule
                \multirow{5}{10pt}{Snow} & 1 & \multirow{5}{10pt}{27.8757 $\pm 0.6371$} & 48.6544 & $\pm 0.1278$ & 20.7787 & $\pm 0.5542$ \\
                & 2 & & 48.8333 & $\pm 0.2453$ & 20.9576 & $\pm 0.5467$ \\
                & 3 & & 49.7223 & $\pm 0.3301$ & 21.8466 & $\pm 0.5147$ \\
                & 4 & & 50.0023 & $\pm 0.4704$ & 22.1266 & $\pm 0.5434$ \\
                & 5 & & 49.9184 & $\pm 0.5976$ & 22.0427 & $\pm 0.5395$ \\
                \midrule
                \multirow{5}{10pt}{Frost} & 1 & \multirow{5}{10pt}{27.8757 $\pm 0.6371$} & 42.2587 & $\pm 1.4448$ & 14.3830 & $\pm 1.1510$ \\
                & 2 & & 46.0172 & $\pm 1.3306$ & 18.1415 & $\pm 1.0910$ \\
\bottomrule
            \end{tabular}
        \end{center}
        \label{tab:results_complete_mtl_corruption_rmftl_color_2}
    \end{minipage}
\end{table}

\begin{table}
    \begin{minipage}{0.49\textwidth}
        \caption{RMFTL shape task error rates under individual ImageNet-C data corruption attacks (1-45) with error bars.}
        \tiny
        \begin{center}
            \begin{tabular}{p{0.08\textwidth}p{0.06\textwidth}p{0.1\textwidth}p{0.04\textwidth}p{0.11\textwidth}p{0.04\textwidth}p{0.13\textwidth}}
                \toprule
                Attacks & Severity & $\epsilon_\text{cl}$ & \multicolumn{2}{l}{$\epsilon_\text{at}$} & \multicolumn{2}{l}{$\Delta_\epsilon$} \\
                \midrule
                \multirow{5}{10pt}{Gaussian Noise} & 1 & \multirow{5}{10pt}{26.8296 $\pm 0.5559$} & 51.2400 & $\pm 0.0086$ & 24.4104 & $\pm 0.5513$ \\
                & 2 & & 51.2466 & $\pm 0.0000$ & 24.4170 & $\pm 0.5559$ \\
                & 3 & & 51.2466 & $\pm 0.0000$ & 24.4170 & $\pm 0.5559$ \\
                & 4 & & 51.2466 & $\pm 0.0000$ & 24.4170 & $\pm 0.5559$ \\
                & 5 & & 51.2466 & $\pm 0.0000$ & 24.4170 & $\pm 0.5559$ \\
                \midrule
                \multirow{5}{10pt}{Shot Noise} & 1 & \multirow{5}{10pt}{26.8296 $\pm 0.5559$} & 50.8648 & $\pm 0.1337$ & 24.0353 & $\pm 0.5123$ \\
                & 2 & & 51.2466 & $\pm 0.0000$ & 24.4170 & $\pm 0.5559$ \\
\bottomrule
            \end{tabular}
        \end{center}
        \label{tab:results_complete_mtl_corruption_rmftl_shape_1}
    \end{minipage}
    \hspace{2pt}
    \begin{minipage}{0.49\textwidth}
        \caption{RMFTL shape task error rates under individual ImageNet-C data corruption attacks (46-95) with error bars.}
        \tiny
        \begin{center}
            \begin{tabular}{p{0.08\textwidth}p{0.06\textwidth}p{0.1\textwidth}p{0.04\textwidth}p{0.11\textwidth}p{0.04\textwidth}p{0.13\textwidth}}
                \toprule
                Attacks & Severity & $\epsilon_\text{cl}$ & \multicolumn{2}{l}{$\epsilon_\text{at}$} & \multicolumn{2}{l}{$\Delta_\epsilon$} \\
                \midrule
                \multirow{5}{10pt}{Snow} & 1 & \multirow{5}{10pt}{26.8296 $\pm 0.5559$} & 47.6908 & $\pm 0.2382$ & 20.8612 & $\pm 0.5367$ \\
                & 2 & & 48.1154 & $\pm 0.1690$ & 21.2859 & $\pm 0.5466$ \\
                & 3 & & 49.4938 & $\pm 0.5195$ & 22.6642 & $\pm 0.5628$ \\
                & 4 & & 50.0182 & $\pm 0.5644$ & 23.1886 & $\pm 0.5562$ \\
                & 5 & & 49.6424 & $\pm 0.9333$ & 22.8128 & $\pm 0.7932$ \\
                \midrule
                \multirow{5}{10pt}{Frost} & 1 & \multirow{5}{10pt}{26.8296 $\pm 0.5559$} & 39.0959 & $\pm 1.1684$ & 12.2663 & $\pm 0.8258$ \\
                & 2 & & 42.1259 & $\pm 0.9717$ & 15.2964 & $\pm 0.6341$ \\
\bottomrule
            \end{tabular}
        \end{center}
        \label{tab:results_complete_mtl_corruption_rmftl_shape_2}
    \end{minipage}
\end{table}

\begin{table}
    \begin{minipage}{0.49\textwidth}
        \caption{RMFTL symbol task error rates under individual ImageNet-C data corruption attacks (1-45) with error bars.}
        \tiny
        \begin{center}
            \begin{tabular}{p{0.08\textwidth}p{0.06\textwidth}p{0.1\textwidth}p{0.04\textwidth}p{0.11\textwidth}p{0.04\textwidth}p{0.13\textwidth}}
                \toprule
                Attacks & Severity & $\epsilon_\text{cl}$ & \multicolumn{2}{l}{$\epsilon_\text{at}$} & \multicolumn{2}{l}{$\Delta_\epsilon$} \\
                \midrule
                \multirow{5}{10pt}{Gaussian Noise} & 1 & \multirow{5}{10pt}{38.9440 $\pm 0.1248$} & 39.2332 & $\pm 0.0000$ & 0.2893 & $\pm 0.1248$ \\
                & 2 & & 39.2332 & $\pm 0.0000$ & 0.2893 & $\pm 0.1248$ \\
                & 3 & & 39.2332 & $\pm 0.0000$ & 0.2893 & $\pm 0.1248$ \\
                & 4 & & 39.2332 & $\pm 0.0000$ & 0.2893 & $\pm 0.1248$ \\
                & 5 & & 39.2332 & $\pm 0.0000$ & 0.2893 & $\pm 0.1248$ \\
                \midrule
                \multirow{5}{10pt}{Shot Noise} & 1 & \multirow{5}{10pt}{38.9440 $\pm 0.1248$} & 39.2332 & $\pm 0.0000$ & 0.2893 & $\pm 0.1248$ \\
                & 2 & & 39.2332 & $\pm 0.0000$ & 0.2893 & $\pm 0.1248$ \\
\bottomrule
            \end{tabular}
        \end{center}
        \label{tab:results_complete_mtl_corruption_rmftl_symbol_1}
    \end{minipage}
    \hspace{2pt}
    \begin{minipage}{0.49\textwidth}
        \caption{RMFTL symbol task error rates under individual ImageNet-C data corruption attacks (46-95) with error bars.}
        \tiny
        \begin{center}
            \begin{tabular}{p{0.08\textwidth}p{0.06\textwidth}p{0.1\textwidth}p{0.04\textwidth}p{0.11\textwidth}p{0.04\textwidth}p{0.13\textwidth}}
                \toprule
                Attacks & Severity & $\epsilon_\text{cl}$ & \multicolumn{2}{l}{$\epsilon_\text{at}$} & \multicolumn{2}{l}{$\Delta_\epsilon$} \\
                \midrule
                \multirow{5}{10pt}{Snow} & 1 & \multirow{5}{10pt}{38.9440 $\pm 0.1248$} & 39.2332 & $\pm 0.0000$ & 0.2893 & $\pm 0.1248$ \\
                & 2 & & 39.2332 & $\pm 0.0000$ & 0.2893 & $\pm 0.1248$ \\
                & 3 & & 39.2332 & $\pm 0.0000$ & 0.2893 & $\pm 0.1248$ \\
                & 4 & & 39.2332 & $\pm 0.0000$ & 0.2893 & $\pm 0.1248$ \\
                & 5 & & 39.2332 & $\pm 0.0000$ & 0.2893 & $\pm 0.1248$ \\
                \midrule
                \multirow{5}{10pt}{Frost} & 1 & \multirow{5}{10pt}{38.9440 $\pm 0.1248$} & 39.2319 & $\pm 0.0026$ & 0.2880 & $\pm 0.1238$ \\
                & 2 & & 39.2332 & $\pm 0.0000$ & 0.2893 & $\pm 0.1248$ \\
\bottomrule
            \end{tabular}
        \end{center}
        \label{tab:results_complete_mtl_corruption_rmftl_symbol_2}
    \end{minipage}
\end{table}

\begin{table}
    \begin{minipage}{0.49\textwidth}
        \caption{RMFTL text task error rates under individual ImageNet-C data corruption attacks (1-45) with error bars.}
        \tiny
        \begin{center}
            \begin{tabular}{p{0.07\textwidth}p{0.05\textwidth}p{0.1\textwidth}p{0.04\textwidth}p{0.11\textwidth}p{0.09\textwidth}p{0.11\textwidth}}
                \toprule
                Attacks & Severity & $\epsilon_\text{cl}$ & \multicolumn{2}{l}{$\epsilon_\text{at}$} & \multicolumn{2}{l}{$\Delta_\epsilon$} \\
                \midrule
                \multirow{5}{10pt}{Gaussian Noise} & 1 & \multirow{5}{10pt}{15.4152 $\pm 0.0681$} & 15.5665 & $\pm 0.0000$ & 0.1512 & $\pm 0.0681$ \\
                & 2 & & 15.5665 & $\pm 0.0000$ & 0.1512 & $\pm 0.0681$ \\
                & 3 & & 15.5665 & $\pm 0.0000$ & 0.1512 & $\pm 0.0681$ \\
                & 4 & & 15.5665 & $\pm 0.0000$ & 0.1512 & $\pm 0.0681$ \\
                & 5 & & 15.5665 & $\pm 0.0000$ & 0.1512 & $\pm 0.0681$ \\
                \midrule
                \multirow{5}{10pt}{Shot Noise} & 1 & \multirow{5}{10pt}{15.4152 $\pm 0.0681$} & 15.5665 & $\pm 0.0000$ & 0.1512 & $\pm 0.0681$ \\
                & 2 & & 15.5665 & $\pm 0.0000$ & 0.1512 & $\pm 0.0681$ \\
\bottomrule
            \end{tabular}
        \end{center}
        \label{tab:results_complete_mtl_corruption_rmftl_text_1}
    \end{minipage}
    \hspace{2pt}
    \begin{minipage}{0.49\textwidth}
        \caption{RMFTL text task error rates under individual ImageNet-C data corruption attacks (46-95) with error bars.}
        \tiny
        \begin{center}
            \begin{tabular}{p{0.08\textwidth}p{0.06\textwidth}p{0.1\textwidth}p{0.04\textwidth}p{0.11\textwidth}p{0.04\textwidth}p{0.13\textwidth}}
                \toprule
                Attacks & Severity & $\epsilon_\text{cl}$ & \multicolumn{2}{l}{$\epsilon_\text{at}$} & \multicolumn{2}{l}{$\Delta_\epsilon$} \\
                \midrule
                \multirow{5}{10pt}{Snow} & 1 & \multirow{5}{10pt}{15.4152 $\pm 0.0681$} & 15.5539 & $\pm 0.0082$ & 0.1387 & $\pm 0.0604$ \\
                & 2 & & 15.5572 & $\pm 0.0053$ & 0.1420 & $\pm 0.0635$ \\
                & 3 & & 15.5652 & $\pm 0.0034$ & 0.1499 & $\pm 0.0676$ \\
                & 4 & & 15.5665 & $\pm 0.0021$ & 0.1512 & $\pm 0.0682$ \\
                & 5 & & 15.5625 & $\pm 0.0082$ & 0.1473 & $\pm 0.0641$ \\
                \midrule
                \multirow{5}{10pt}{Frost} & 1 & \multirow{5}{10pt}{15.4152 $\pm 0.0681$} & 15.5308 & $\pm 0.0218$ & 0.1156 & $\pm 0.0501$ \\
                & 2 & & 15.5592 & $\pm 0.0109$ & 0.1440 & $\pm 0.0587$ \\
\bottomrule
            \end{tabular}
        \end{center}
        \label{tab:results_complete_mtl_corruption_rmftl_text_2}
    \end{minipage}
\end{table}

\begin{table}
    \begin{minipage}{0.48\textwidth}
        \caption{RMFTL cumulative relative errors under individual ImageNet-C data corruption attacks (1-45) with error bars.}
        \scriptsize
        \begin{center}
            \begin{tabular}{llll}
                \toprule
                Attacks & Severity & \multicolumn{2}{l}{$\epsilon_\text{cu}$} \\
                \midrule
                \multirow{5}{10pt}{Gaussian Noise} & 1 & 48.2152 & $\pm 1.3794$ \\
                & 2 & 48.2284 & $\pm 1.3823$ \\
                & 3 & 48.2284 & $\pm 1.3823$ \\
                & 4 & 48.2284 & $\pm 1.3823$ \\
                & 5 & 48.2284 & $\pm 1.3823$ \\
                \midrule
                \multirow{5}{10pt}{Shot Noise} & 1 & 47.7615 & $\pm 1.3265$ \\
                & 2 & 48.2284 & $\pm 1.3823$ \\
\bottomrule
            \end{tabular}
        \end{center}
        \label{tab:results_complete_mtl_corruption_rmftl_cumulative_1}
    \end{minipage}
    \quad
    \begin{minipage}{0.48\textwidth}
        \caption{RMFTL cumulative relative errors under individual ImageNet-C data corruption attacks (46-95) with error bars.}
        \scriptsize
        \begin{center}
            \begin{tabular}{llll}
                \toprule
                Attacks & Severity & \multicolumn{2}{l}{$\epsilon_\text{cu}$} \\
                \midrule
                \multirow{5}{10pt}{Snow} & 1 & 42.0678 & $\pm 1.2421$ \\
                & 2 & 42.6748 & $\pm 1.2688$ \\
                & 3 & 44.9500 & $\pm 1.1844$ \\
                & 4 & 45.7557 & $\pm 1.1881$ \\
                & 5 & 45.2921 & $\pm 1.3492$ \\
                \midrule
                \multirow{5}{10pt}{Frost} & 1 & 27.0528 & $\pm 1.8843$ \\
                & 2 & 33.8711 & $\pm 1.6582$ \\
\bottomrule
            \end{tabular}
        \end{center}
        \label{tab:results_complete_mtl_corruption_rmftl_cumulative_2}
    \end{minipage}
\end{table}

\begin{table}
    \begin{minipage}{0.49\textwidth}
        \caption{VMFTL color task error rates under individual ImageNet-C data corruption attacks (1-45) with error bars.}
        \tiny
        \begin{center}
            \begin{tabular}{p{0.08\textwidth}p{0.06\textwidth}p{0.1\textwidth}p{0.04\textwidth}p{0.11\textwidth}p{0.04\textwidth}p{0.13\textwidth}}
                \toprule
                Attacks & Severity & $\epsilon_\text{cl}$ & \multicolumn{2}{l}{$\epsilon_\text{at}$} & \multicolumn{2}{l}{$\Delta_\epsilon$} \\
                \midrule
                \multirow{5}{10pt}{Gaussian Noise} & 1 & \multirow{5}{10pt}{19.7946 $\pm 0.0768$} & 26.5482 & $\pm 0.1585$ & 6.7536 & $\pm 0.1729$ \\
                & 2 & & 29.2626 & $\pm 0.2105$ & 9.4680 & $\pm 0.2380$ \\
                & 3 & & 33.3395 & $\pm 0.2709$ & 13.5449 & $\pm 0.2892$ \\
                & 4 & & 37.3332 & $\pm 0.3188$ & 17.5386 & $\pm 0.3524$ \\
                & 5 & & 42.3168 & $\pm 0.4259$ & 22.5222 & $\pm 0.4585$ \\
                \midrule
                \multirow{5}{10pt}{Shot Noise} & 1 & \multirow{5}{10pt}{19.7946 $\pm 0.0768$} & 26.5687 & $\pm 0.1788$ & 6.7741 & $\pm 0.2086$ \\
                & 2 & & 29.4575 & $\pm 0.2283$ & 9.6628 & $\pm 0.2670$ \\
\bottomrule
            \end{tabular}
        \end{center}
        \label{tab:results_complete_mtl_corruption_vmftl_color_1}
    \end{minipage}
    \hspace{2pt}
    \begin{minipage}{0.49\textwidth}
        \caption{VMFTL color task error rates under individual ImageNet-C data corruption attacks (46-95) with error bars.}
        \tiny
        \begin{center}
            \begin{tabular}{p{0.08\textwidth}p{0.06\textwidth}p{0.1\textwidth}p{0.04\textwidth}p{0.11\textwidth}p{0.04\textwidth}p{0.13\textwidth}}
                \toprule
                Attacks & Severity & $\epsilon_\text{cl}$ & \multicolumn{2}{l}{$\epsilon_\text{at}$} & \multicolumn{2}{l}{$\Delta_\epsilon$} \\
                \midrule
                \multirow{5}{10pt}{Snow} & 1 & \multirow{5}{10pt}{19.7946 $\pm 0.0768$} & 31.9499 & $\pm 1.6342$ & 12.1553 & $\pm 1.5828$ \\
                & 2 & & 38.0451 & $\pm 1.2407$ & 18.2505 & $\pm 1.1933$ \\
                & 3 & & 42.7963 & $\pm 1.0973$ & 23.0017 & $\pm 1.0452$ \\
                & 4 & & 46.3877 & $\pm 0.2111$ & 26.5931 & $\pm 0.1809$ \\
                & 5 & & 45.6097 & $\pm 2.0103$ & 25.8151 & $\pm 1.9539$ \\
                \midrule
                \multirow{5}{10pt}{Frost} & 1 & \multirow{5}{10pt}{19.7946 $\pm 0.0768$} & 24.8932 & $\pm 2.4806$ & 5.0986 & $\pm 2.5397$ \\
                & 2 & & 29.2633 & $\pm 4.2636$ & 9.4687 & $\pm 4.3245$ \\
\bottomrule
            \end{tabular}
        \end{center}
        \label{tab:results_complete_mtl_corruption_vmftl_color_2}
    \end{minipage}
\end{table}

\begin{table}
    \begin{minipage}{0.49\textwidth}
        \caption{VMFTL shape task error rates under individual ImageNet-C data corruption attacks (1-45) with error bars.}
        \tiny
        \begin{center}
            \begin{tabular}{p{0.08\textwidth}p{0.06\textwidth}p{0.1\textwidth}p{0.04\textwidth}p{0.11\textwidth}p{0.04\textwidth}p{0.13\textwidth}}
                \toprule
                Attacks & Severity & $\epsilon_\text{cl}$ & \multicolumn{2}{l}{$\epsilon_\text{at}$} & \multicolumn{2}{l}{$\Delta_\epsilon$} \\
                \midrule
                \multirow{5}{10pt}{Gaussian Noise} & 1 & \multirow{5}{10pt}{19.1573 $\pm 0.0544$} & 23.9223 & $\pm 0.3317$ & 4.7650 & $\pm 0.2905$ \\
                & 2 & & 26.5311 & $\pm 0.3931$ & 7.3738 & $\pm 0.3517$ \\
                & 3 & & 30.4138 & $\pm 0.4264$ & 11.2565 & $\pm 0.3917$ \\
                & 4 & & 34.2007 & $\pm 0.5153$ & 15.0434 & $\pm 0.4815$ \\
                & 5 & & 38.7610 & $\pm 0.3819$ & 19.6037 & $\pm 0.3725$ \\
                \midrule
                \multirow{5}{10pt}{Shot Noise} & 1 & \multirow{5}{10pt}{19.1573 $\pm 0.0544$} & 24.1819 & $\pm 0.0867$ & 5.0246 & $\pm 0.1093$ \\
                & 2 & & 26.9432 & $\pm 0.1699$ & 7.7859 & $\pm 0.1317$ \\
\bottomrule
            \end{tabular}
        \end{center}
        \label{tab:results_complete_mtl_corruption_vmftl_shape_1}
    \end{minipage}
    \hspace{2pt}
    \begin{minipage}{0.49\textwidth}
        \caption{VMFTL shape task error rates under individual ImageNet-C data corruption attacks (46-95) with error bars.}
        \tiny
        \begin{center}
            \begin{tabular}{p{0.08\textwidth}p{0.06\textwidth}p{0.1\textwidth}p{0.04\textwidth}p{0.11\textwidth}p{0.04\textwidth}p{0.13\textwidth}}
                \toprule
                Attacks & Severity & $\epsilon_\text{cl}$ & \multicolumn{2}{l}{$\epsilon_\text{at}$} & \multicolumn{2}{l}{$\Delta_\epsilon$} \\
                \midrule
                \multirow{5}{10pt}{Snow} & 1 & \multirow{5}{10pt}{19.1573 $\pm 0.0544$} & 30.3880 & $\pm 1.2423$ & 11.2307 & $\pm 1.2775$ \\
                & 2 & & 36.9415 & $\pm 0.5800$ & 17.7842 & $\pm 0.6210$ \\
                & 3 & & 41.3433 & $\pm 1.0593$ & 22.1860 & $\pm 1.0860$ \\
                & 4 & & 45.1336 & $\pm 0.4827$ & 25.9763 & $\pm 0.4976$ \\
                & 5 & & 44.4064 & $\pm 2.4560$ & 25.2491 & $\pm 2.4768$ \\
                \midrule
                \multirow{5}{10pt}{Frost} & 1 & \multirow{5}{10pt}{19.1573 $\pm 0.0544$} & 23.9454 & $\pm 2.3614$ & 4.7882 & $\pm 2.3526$ \\
                & 2 & & 28.6253 & $\pm 4.3904$ & 9.4680 & $\pm 4.3740$ \\
\bottomrule
            \end{tabular}
        \end{center}
        \label{tab:results_complete_mtl_corruption_vmftl_shape_2}
    \end{minipage}
\end{table}
        
\begin{table}
    \begin{minipage}{0.49\textwidth}
        \caption{VMFTL symbol task error rates under individual ImageNet-C data corruption attacks (1-45) with error bars.}
        \tiny
        \begin{center}
            \begin{tabular}{p{0.07\textwidth}p{0.06\textwidth}p{0.09\textwidth}p{0.04\textwidth}p{0.11\textwidth}p{0.09\textwidth}p{0.11\textwidth}}
                \toprule
                Attacks & Severity & $\epsilon_\text{cl}$ & \multicolumn{2}{l}{$\epsilon_\text{at}$} & \multicolumn{2}{l}{$\Delta_\epsilon$} \\
                \midrule
                \multirow{5}{10pt}{Gaussian Noise} & 1 & \multirow{5}{10pt}{31.8588 $\pm 0.0857$} & 33.0938 & $\pm 0.0563$ & 1.2350 & $\pm 0.0742$ \\
                & 2 & & 33.9332 & $\pm 0.1209$ & 2.0744 & $\pm 0.1329$ \\
                & 3 & & 35.1332 & $\pm 0.1256$ & 3.2744 & $\pm 0.1277$ \\
                & 4 & & 36.0539 & $\pm 0.1930$ & 4.1951 & $\pm 0.1819$ \\
                & 5 & & 37.5009 & $\pm 0.2044$ & 5.6421 & $\pm 0.2541$ \\
                \midrule
                \multirow{5}{10pt}{Shot Noise} & 1 & \multirow{5}{10pt}{31.8588 $\pm 0.0857$} & 32.9452 & $\pm 0.0785$ & 1.0864 & $\pm 0.0993$ \\
                & 2 & & 33.7926 & $\pm 0.1074$ & 1.9338 & $\pm 0.1179$ \\
\bottomrule
            \end{tabular}
        \end{center}
        \label{tab:results_complete_mtl_corruption_vmftl_symbol_1}
    \end{minipage}
    \hspace{2pt}
    \begin{minipage}{0.49\textwidth}
        \caption{VMFTL symbol task error rates under individual ImageNet-C data corruption attacks (46-95) with error bars.}
        \tiny
        \begin{center}
            \begin{tabular}{p{0.078\textwidth}p{0.045\textwidth}p{0.09\textwidth}p{0.038\textwidth}p{0.11\textwidth}p{0.09\textwidth}p{0.11\textwidth}}
                \toprule
                Attacks & Severity & $\epsilon_\text{cl}$ & \multicolumn{2}{l}{$\epsilon_\text{at}$} & \multicolumn{2}{l}{$\Delta_\epsilon$} \\
                \midrule
                \multirow{5}{10pt}{Snow} & 1 & \multirow{5}{10pt}{31.8588 $\pm 0.0857$} & 35.4978 & $\pm 0.2813$ & 3.6390 & $\pm 0.2338$ \\
                & 2 & & 37.0637 & $\pm 0.1895$ & 5.2049 & $\pm 0.1438$ \\
                & 3 & & 37.5082 & $\pm 0.2431$ & 5.6494 & $\pm 0.1908$ \\
                & 4 & & 38.3595 & $\pm 0.1375$ & 6.5007 & $\pm 0.1312$ \\
                & 5 & & 38.3674 & $\pm 0.2372$ & 6.5086 & $\pm 0.2797$ \\
                \midrule
                \multirow{5}{10pt}{Frost} & 1 & \multirow{5}{10pt}{31.8588 $\pm 0.0857$} & 33.1988 & $\pm 0.5826$ & 1.3400 & $\pm 0.6088$ \\
                & 2 & & 34.4834 & $\pm 1.1109$ & 2.6246 & $\pm 1.1324$ \\
\bottomrule
            \end{tabular}
        \end{center}
        \label{tab:results_complete_mtl_corruption_vmftl_symbol_2}
    \end{minipage}
\end{table}

\begin{table}
    \begin{minipage}{0.49\textwidth}
        \caption{VMFTL text task error rates under individual ImageNet-C data corruption attacks (1-45) with error bars.}
        \tiny
        \begin{center}
            \begin{tabular}{p{0.07\textwidth}p{0.05\textwidth}p{0.1\textwidth}p{0.04\textwidth}p{0.11\textwidth}p{0.090\textwidth}p{0.11\textwidth}}
                \toprule
                Attacks & Severity & $\epsilon_\text{cl}$ & \multicolumn{2}{l}{$\epsilon_\text{at}$} & \multicolumn{2}{l}{$\Delta_\epsilon$} \\
                \midrule
                \multirow{5}{10pt}{Gaussian Noise} & 1 & \multirow{5}{10pt}{12.7570 $\pm 0.0238$} & 13.8104 & $\pm 0.0671$ & 1.0534 & $\pm 0.0708$ \\
                & 2 & & 14.2284 & $\pm 0.1038$ & 1.4715 & $\pm 0.0911$ \\
                & 3 & & 14.6663 & $\pm 0.0907$ & 1.9093 & $\pm 0.0875$ \\
                & 4 & & 15.0725 & $\pm 0.0973$ & 2.3155 & $\pm 0.0964$ \\
                & 5 & & 15.4522 & $\pm 0.0721$ & 2.6952 & $\pm 0.0726$ \\
                \midrule
                \multirow{5}{10pt}{Shot Noise} & 1 & \multirow{5}{10pt}{12.7570 $\pm 0.0238$} & 13.4643 & $\pm 0.0463$ & 0.7073 & $\pm 0.0480$ \\
                & 2 & & 14.0052 & $\pm 0.0943$ & 1.2482 & $\pm 0.0871$ \\
\bottomrule
            \end{tabular}
        \end{center}
        \label{tab:results_complete_mtl_corruption_vmftl_text_1}
    \end{minipage}
    \hspace{2pt}
    \begin{minipage}{0.49\textwidth}
        \caption{VMFTL text task error rates under individual ImageNet-C data corruption attacks (46-95) with error bars.}
        \tiny
        \begin{center}
            \begin{tabular}{p{0.08\textwidth}p{0.06\textwidth}p{0.1\textwidth}p{0.04\textwidth}p{0.11\textwidth}p{0.04\textwidth}p{0.13\textwidth}}
                \toprule
                Attacks & Severity & $\epsilon_\text{cl}$ & \multicolumn{2}{l}{$\epsilon_\text{at}$} & \multicolumn{2}{l}{$\Delta_\epsilon$} \\
                \midrule
                \multirow{5}{10pt}{Snow} & 1 & \multirow{5}{10pt}{12.7570 $\pm 0.0238$} & 14.5666 & $\pm 0.0904$ & 1.8096 & $\pm 0.0813$ \\
                & 2 & & 15.1775 & $\pm 0.2405$ & 2.4205 & $\pm 0.2377$ \\
                & 3 & & 15.3631 & $\pm 0.1216$ & 2.6061 & $\pm 0.1209$ \\
                & 4 & & 15.6088 & $\pm 0.2348$ & 2.8518 & $\pm 0.2294$ \\
                & 5 & & 15.9007 & $\pm 0.3653$ & 3.1437 & $\pm 0.3565$ \\
                \midrule
                \multirow{5}{10pt}{Frost} & 1 & \multirow{5}{10pt}{12.7570 $\pm 0.0238$} & 13.4538 & $\pm 0.3936$ & 0.6968 & $\pm 0.3849$ \\
                & 2 & & 14.0382 & $\pm 0.5812$ & 1.2812 & $\pm 0.5697$ \\
\bottomrule
            \end{tabular}
        \end{center}
        \label{tab:results_complete_mtl_corruption_vmftl_text_2}
    \end{minipage}
\end{table}

\begin{table}
    \begin{minipage}{0.48\textwidth}
        \caption{VMFTL cumulative relative errors under individual ImageNet-C data corruption attacks (1-45) with error bars.}
        \scriptsize
        \begin{center}
            \begin{tabular}{llll}
                \toprule
                Attacks & Severity & \multicolumn{2}{l}{$\epsilon_\text{cu}$} \\
                \midrule
                \multirow{5}{10pt}{Gaussian Noise} & 1 & 13.8071 & $\pm 0.1678$ \\
                & 2 & 20.3877 & $\pm 0.2754$ \\
                & 3 & 29.9851 & $\pm 0.2231$ \\
                & 4 & 39.0926 & $\pm 0.3608$ \\
                & 5 & 50.4633 & $\pm 0.3520$ \\
                \midrule
                \multirow{5}{10pt}{Shot Noise} & 1 & 13.5924 & $\pm 0.3395$ \\
                & 2 & 20.6307 & $\pm 0.3005$ \\
\bottomrule
            \end{tabular}
        \end{center}
        \label{tab:results_complete_mtl_corruption_vmftl_cumulative_1}
    \end{minipage}
    \quad
    \begin{minipage}{0.48\textwidth}
        \caption{VMFTL cumulative relative errors under individual ImageNet-C data corruption attacks (46-95) with error bars.}
        \scriptsize
        \begin{center}
            \begin{tabular}{llll}
                \toprule
                Attacks & Severity & \multicolumn{2}{l}{$\epsilon_\text{cu}$} \\
                \midrule
                \multirow{5}{10pt}{Snow} & 1 & 28.8347 & $\pm 2.8399$ \\
                & 2 & 43.6601 & $\pm 1.5265$ \\
                & 3 & 53.4432 & $\pm 2.1954$ \\
                & 4 & 61.9219 & $\pm 0.4033$ \\
                & 5 & 60.7166 & $\pm 4.3581$ \\
                \midrule
                \multirow{5}{10pt}{Frost} & 1 & 11.9235 & $\pm 5.8552$ \\
                & 2 & 22.8425 & $\pm 10.3898$ \\
\bottomrule
            \end{tabular}
        \end{center}
        \label{tab:results_complete_mtl_corruption_vmftl_cumulative_2}
    \end{minipage}
\end{table}

\begin{table}
    \vspace{48pt}
    \caption{MTL model color task error rates under color quantization attacks.}
    \scriptsize
    \begin{center}
        \begin{tabular}{lllllll}
            \toprule
            Models & $\epsilon_\text{cl}$ & & $\epsilon_\text{at}$ & & $\Delta_\epsilon$ & \\
            \midrule
            RMFT & 8.8584 & $\pm 0.1480$ & 32.9029 & $\pm 0.5508$ & 24.0445 & $\pm 0.4939$ \\
            \midrule
            VMFT & 7.6914 & $\pm 0.1164$ & 31.7571 & $\pm 0.5390$ & 24.0656 & $\pm 0.5941$ \\
            \midrule
            RMFTL & 27.8757 & $\pm 0.7123$ & 46.4881 & $\pm 0.4789$ & 18.6124 & $\pm 0.3063$ \\
            \midrule
            VMFTL & 19.7946 & $\pm 0.0858$ & 36.3049 & $\pm 0.1351$ & 16.5103 & $\pm 0.1667$ \\
            \bottomrule
        \end{tabular}
    \end{center}
    \label{tab:results_complete_mtl_color_quantization_color}
\end{table}

\begin{table}
    \caption{MTL model shape task error rates under color quantization attacks.}
    \scriptsize
    \begin{center}
        \begin{tabular}{lllllll}
            \toprule
            Models & $\epsilon_\text{cl}$ & & $\epsilon_\text{at}$ & & $\Delta_\epsilon$ & \\
            \midrule
            RMFT & 8.5698 & $\pm 0.1706$ & 26.5720 & $\pm 0.4991$ & 18.0022 & $\pm 0.4347$ \\
            \midrule
            VMFT & 7.5950 & $\pm 0.0853$ & 26.2940 & $\pm 0.5089$ & 18.6989 & $\pm 0.5637$ \\
            \midrule
            RMFTL & 26.8296 & $\pm 0.6215$ & 37.2598 & $\pm 0.3852$ & 10.4303 & $\pm 0.2691$ \\
            \midrule
            VMFTL & 19.1573 & $\pm 0.0608$ & 27.2417 & $\pm 0.0506$ & 8.0844 & $\pm 0.0797$ \\
            \bottomrule
        \end{tabular}
    \end{center}
    \label{tab:results_complete_mtl_color_quantization_shape}
\end{table}

\begin{table}
    \caption{MTL model symbol task error rates under color quantization attacks.}
    \scriptsize
    \begin{center}
        \begin{tabular}{lllllll}
            \toprule
            Models & $\epsilon_\text{cl}$ & & $\epsilon_\text{at}$ & & $\Delta_\epsilon$ & \\
            \midrule
            RMFT & 11.9684 & $\pm 0.3968$ & 26.7814 & $\pm 0.1698$ & 14.8129 & $\pm 0.2357$ \\
            \midrule
            VMFT & 11.0768 & $\pm 0.2277$ & 26.3118 & $\pm 0.4633$ & 15.2350 & $\pm 0.5113$ \\
            \midrule
            RMFTL & 38.9440 & $\pm 0.1396$ & 39.2035 & $\pm 0.0279$ & 0.2596 & $\pm 0.1150$ \\
            \midrule
            VMFTL & 31.8588 & $\pm 0.0958$ & 35.8723 & $\pm 0.0411$ & 4.0135 & $\pm 0.1088$ \\
            \bottomrule
        \end{tabular}
    \end{center}
    \label{tab:results_complete_mtl_color_quantization_symbol}
\end{table}

\begin{table}
    \caption{MTL model text task error rates under color quantization attacks.}
    \scriptsize
    \begin{center}
        \begin{tabular}{lllllll}
            \toprule
            Models & $\epsilon_\text{cl}$ & & $\epsilon_\text{at}$ & & $\Delta_\epsilon$ & \\
            \midrule
            RMFT & 4.3926 & $\pm 0.1334$ & 12.6044 & $\pm 0.1302$ & 8.2119 & $\pm 0.1249$ \\
            \midrule
            VMFT & 3.9970 & $\pm 0.1592$ & 12.0913 & $\pm 0.1010$ & 8.0943 & $\pm 0.1572$ \\
            \midrule
            RMFTL & 15.4152 & $\pm 0.0761$ & 15.5665 & $\pm 0.0023$ & 0.1512 & $\pm 0.0763$ \\
            \midrule
            VMFTL & 12.7570 & $\pm 0.0267$ & 14.7284 & $\pm 0.0369$ & 1.9714 & $\pm 0.0571$ \\
            \bottomrule
        \end{tabular}
    \end{center}
    \label{tab:results_complete_mtl_color_quantization_text}
\end{table}

\begin{table}
    \caption{MTL model color task cumulative relative errors and relative error correlations under color quantization attacks.}
    \scriptsize
    \begin{center}
        %
    \end{center}
    \label{tab:results_complete_mtl_color_quantization_cumulative}
\end{table}


\begin{thebibliography}{10}

\bibitem{aghdam2015unified}
Hamed~Habibi Aghdam, Elnaz~Jahani Heravi, and Domenec Puig.
\newblock A unified framework for coarse-to-fine recognition of traffic signs using bayesian network and visual attributes.
\newblock In {\em VISAPP (2)}, pages 87--96, 2015.

\bibitem{alam2023medic}
Firoj Alam, Tanvirul Alam, Md~Arid Hasan, Abul Hasnat, Muhammad Imran, and Ferda Ofli.
\newblock Medic: a multi-task learning dataset for disaster image classification.
\newblock {\em Neural Computing and Applications}, 35(3):2609--2632, 2023.

\bibitem{armitage2020mlm}
Jason Armitage, Endri Kacupaj, Golsa Tahmasebzadeh, Swati, Maria Maleshkova, Ralph Ewerth, and Jens Lehmann.
\newblock Mlm: A benchmark dataset for multitask learning with multiple languages and modalities.
\newblock In {\em Proceedings of the 29th ACM International Conference on Information \& Knowledge Management}, pages 2967--2974, 2020.

\bibitem{beery2018recognition}
Sara Beery, Grant Van~Horn, and Pietro Perona.
\newblock Recognition in terra incognita.
\newblock In {\em Proceedings of the European conference on computer vision (ECCV)}, pages 456--473, 2018.

\bibitem{berghoff2021robustness}
Christian Berghoff, Pavol Bielik, Matthias Neu, Petar Tsankov, and Arndt Von~Twickel.
\newblock Robustness testing of ai systems: a case study for traffic sign recognition.
\newblock In {\em Artificial Intelligence Applications and Innovations: 17th IFIP WG 12.5 International Conference, AIAI 2021, Hersonissos, Crete, Greece, June 25--27, 2021, Proceedings 17}, pages 256--267. Springer, 2021.

\bibitem{bochkovskiy2020yolov4}
Alexey Bochkovskiy, Chien-Yao Wang, and Hong-Yuan~Mark Liao.
\newblock Yolov4: Optimal speed and accuracy of object detection.
\newblock {\em arXiv preprint arXiv:2004.10934}, 2020.

\bibitem{caruana1997multitask}
Rich Caruana.
\newblock Multitask learning.
\newblock {\em Machine learning}, 28:41--75, 1997.

\bibitem{croce2020robustbench}
Francesco Croce, Maksym Andriushchenko, Vikash Sehwag, Edoardo Debenedetti, Nicolas Flammarion, Mung Chiang, Prateek Mittal, and Matthias Hein.
\newblock Robustbench: a standardized adversarial robustness benchmark.
\newblock {\em arXiv preprint arXiv:2010.09670}, 2020.

\bibitem{croce2020reliable}
Francesco Croce and Matthias Hein.
\newblock Reliable evaluation of adversarial robustness with an ensemble of diverse parameter-free attacks.
\newblock In {\em International conference on machine learning}, pages 2206--2216. PMLR, 2020.

\bibitem{dosovitskiy2020image}
Alexey Dosovitskiy, Lucas Beyer, Alexander Kolesnikov, Dirk Weissenborn, Xiaohua Zhai, Thomas Unterthiner, Mostafa Dehghani, Matthias Minderer, Georg Heigold, Sylvain Gelly, et~al.
\newblock An image is worth 16x16 words: Transformers for image recognition at scale.
\newblock {\em arXiv preprint arXiv:2010.11929}, 2020.

\bibitem{eger2020hero}
Steffen Eger and Yannik Benz.
\newblock From hero to z$\backslash$'eroe: A benchmark of low-level adversarial attacks.
\newblock {\em arXiv preprint arXiv:2010.05648}, 2020.

\bibitem{ertler2020mapillary}
Christian Ertler, Jerneja Mislej, Tobias Ollmann, Lorenzo Porzi, Gerhard Neuhold, and Yubin Kuang.
\newblock The mapillary traffic sign dataset for detection and classification on a global scale.
\newblock In {\em Computer Vision--ECCV 2020: 16th European Conference, Glasgow, UK, August 23--28, 2020, Proceedings, Part XXIII 16}, pages 68--84. Springer, 2020.

\bibitem{girshick2015fast}
Ross Girshick.
\newblock Fast r-cnn.
\newblock In {\em Proceedings of the IEEE international conference on computer vision}, pages 1440--1448, 2015.

\bibitem{girshick2014rich}
Ross Girshick, Jeff Donahue, Trevor Darrell, and Jitendra Malik.
\newblock Rich feature hierarchies for accurate object detection and semantic segmentation.
\newblock In {\em Proceedings of the IEEE conference on computer vision and pattern recognition}, pages 580--587, 2014.

\bibitem{goel2018smartbox}
Akhil Goel, Anirudh Singh, Akshay Agarwal, Mayank Vatsa, and Richa Singh.
\newblock Smartbox: Benchmarking adversarial detection and mitigation algorithms for face recognition.
\newblock In {\em 2018 IEEE 9th international conference on biometrics theory, applications and systems (BTAS)}, pages 1--7. IEEE, 2018.

\bibitem{gurel2021knowledge}
Nezihe~Merve G{\"u}rel, Xiangyu Qi, Luka Rimanic, Ce~Zhang, and Bo~Li.
\newblock Knowledge enhanced machine learning pipeline against diverse adversarial attacks.
\newblock In {\em International Conference on Machine Learning}, pages 3976--3987. PMLR, 2021.

\bibitem{hashemi2022improving}
Atiye~Sadat Hashemi, Saeed Mozaffari, and Shahpour Alirezaee.
\newblock Improving adversarial robustness of traffic sign image recognition networks.
\newblock {\em Displays}, 74:102277, 2022.

\bibitem{he2017mask}
Kaiming He, Georgia Gkioxari, Piotr Doll{\'a}r, and Ross Girshick.
\newblock Mask r-cnn.
\newblock In {\em Proceedings of the IEEE international conference on computer vision}, pages 2961--2969, 2017.

\bibitem{he2016deep}
Kaiming He, Xiangyu Zhang, Shaoqing Ren, and Jian Sun.
\newblock Deep residual learning for image recognition.
\newblock In {\em Proceedings of the IEEE conference on computer vision and pattern recognition}, pages 770--778, 2016.

\bibitem{he2016identity}
Kaiming He, Xiangyu Zhang, Shaoqing Ren, and Jian Sun.
\newblock Identity mappings in deep residual networks.
\newblock In {\em Computer Vision--ECCV 2016: 14th European Conference, Amsterdam, The Netherlands, October 11--14, 2016, Proceedings, Part IV 14}, pages 630--645. Springer, 2016.

\bibitem{hendrycks2019benchmarking}
Dan Hendrycks and Thomas Dietterich.
\newblock Benchmarking neural network robustness to common corruptions and perturbations.
\newblock {\em arXiv preprint arXiv:1903.12261}, 2019.

\bibitem{hingun2022reap}
Nabeel Hingun, Chawin Sitawarin, Jerry Li, and David Wagner.
\newblock Reap: A large-scale realistic adversarial patch benchmark.
\newblock {\em arXiv preprint arXiv:2212.05680}, 2022.

\bibitem{hu2022improving}
Ziniu Hu, Zhe Zhao, Xinyang Yi, Tiansheng Yao, Lichan Hong, Yizhou Sun, and Ed~Chi.
\newblock Improving multi-task generalization via regularizing spurious correlation.
\newblock {\em Advances in Neural Information Processing Systems}, 35:11450--11466, 2022.

\bibitem{huang2017densely}
Gao Huang, Zhuang Liu, Laurens Van Der~Maaten, and Kilian~Q Weinberger.
\newblock Densely connected convolutional networks.
\newblock In {\em Proceedings of the IEEE conference on computer vision and pattern recognition}, pages 4700--4708, 2017.

\bibitem{khosravian2022enhancing}
Amir Khosravian, Abdollah Amirkhani, and Masoud Masih-Tehrani.
\newblock Enhancing the robustness of the convolutional neural networks for traffic sign detection.
\newblock {\em Proceedings of the Institution of Mechanical Engineers, Part D: Journal of Automobile Engineering}, 236(8):1849--1861, 2022.

\bibitem{krizhevsky2017imagenet}
Alex Krizhevsky, Ilya Sutskever, and Geoffrey~E Hinton.
\newblock Imagenet classification with deep convolutional neural networks.
\newblock {\em Communications of the ACM}, 60(6):84--90, 2017.

\bibitem{larsson2011using}
Fredrik Larsson and Michael Felsberg.
\newblock Using fourier descriptors and spatial models for traffic sign recognition.
\newblock In {\em Image Analysis: 17th Scandinavian Conference, SCIA 2011, Ystad, Sweden, May 2011. Proceedings 17}, pages 238--249. Springer, 2011.

\bibitem{larsson2011correlating}
Fredrik Larsson, Michael Felsberg, and P-E Forssen.
\newblock Correlating fourier descriptors of local patches for road sign recognition.
\newblock {\em IET Computer Vision}, 5(4):244--254, 2011.

\bibitem{lin2017focal}
Tsung-Yi Lin, Priya Goyal, Ross Girshick, Kaiming He, and Piotr Doll{\'a}r.
\newblock Focal loss for dense object detection.
\newblock In {\em Proceedings of the IEEE international conference on computer vision}, pages 2980--2988, 2017.

\bibitem{lu2016traffic}
Xiao Lu, Yaonan Wang, Xuanyu Zhou, Zhenjun Zhang, and Zhigang Ling.
\newblock Traffic sign recognition via multi-modal tree-structure embedded multi-task learning.
\newblock {\em IEEE Transactions on Intelligent Transportation Systems}, 18(4):960--972, 2016.

\bibitem{madry2017towards}
Aleksander Madry, Aleksandar Makelov, Ludwig Schmidt, Dimitris Tsipras, and Adrian Vladu.
\newblock Towards deep learning models resistant to adversarial attacks.
\newblock {\em arXiv preprint arXiv:1706.06083}, 2017.

\bibitem{mapillary2022mapillary}
Mapillary.
\newblock Mapillary traffic sign dataset.
\newblock https://www.mapillary.com/dataset/trafficsign, 2022.

\bibitem{mathias2013traffic}
Markus Mathias, Radu Timofte, Rodrigo Benenson, and Luc Van~Gool.
\newblock Traffic sign recognition—how far are we from the solution?
\newblock In {\em The 2013 international joint conference on Neural networks (IJCNN)}, pages 1--8. IEEE, 2013.

\bibitem{michaelis2019benchmarking}
Claudio Michaelis, Benjamin Mitzkus, Robert Geirhos, Evgenia Rusak, Oliver Bringmann, Alexander~S Ecker, Matthias Bethge, and Wieland Brendel.
\newblock Benchmarking robustness in object detection: Autonomous driving when winter is coming.
\newblock {\em arXiv preprint arXiv:1907.07484}, 2019.

\bibitem{mintun2021interaction}
Eric Mintun, Alexander Kirillov, and Saining Xie.
\newblock On interaction between augmentations and corruptions in natural corruption robustness.
\newblock {\em Advances in Neural Information Processing Systems}, 34:3571--3583, 2021.

\bibitem{mogelmose2012vision}
Andreas Mogelmose, Mohan~Manubhai Trivedi, and Thomas~B Moeslund.
\newblock Vision-based traffic sign detection and analysis for intelligent driver assistance systems: Perspectives and survey.
\newblock {\em IEEE Transactions on Intelligent Transportation Systems}, 13(4):1484--1497, 2012.

\bibitem{qian2016traffic}
Rong-Qiang Qian, Yong Yue, Frans Coenen, and Bai-Ling Zhang.
\newblock Traffic sign recognition using visual attribute learning and convolutional neural network.
\newblock In {\em 2016 International Conference on Machine Learning and Cybernetics (ICMLC)}, volume~1, pages 386--391. IEEE, 2016.

\bibitem{redmon2016you}
Joseph Redmon, Santosh Divvala, Ross Girshick, and Ali Farhadi.
\newblock You only look once: Unified, real-time object detection.
\newblock In {\em Proceedings of the IEEE conference on computer vision and pattern recognition}, pages 779--788, 2016.

\bibitem{redmon2017yolo9000}
Joseph Redmon and Ali Farhadi.
\newblock Yolo9000: better, faster, stronger.
\newblock In {\em Proceedings of the IEEE conference on computer vision and pattern recognition}, pages 7263--7271, 2017.

\bibitem{redmon2018yolov3}
Joseph Redmon and Ali Farhadi.
\newblock Yolov3: An incremental improvement.
\newblock {\em arXiv preprint arXiv:1804.02767}, 2018.

\bibitem{ren2015faster}
Shaoqing Ren, Kaiming He, Ross Girshick, and Jian Sun.
\newblock Faster r-cnn: Towards real-time object detection with region proposal networks.
\newblock {\em Advances in neural information processing systems}, 28, 2015.

\bibitem{santurkar2020breeds}
Shibani Santurkar, Dimitris Tsipras, and Aleksander Madry.
\newblock Breeds: Benchmarks for subpopulation shift.
\newblock {\em arXiv preprint arXiv:2008.04859}, 2020.

\bibitem{shakhuro2016russian}
Vladislav~Igorevich Shakhuro and AS~Konouchine.
\newblock Russian traffic sign images dataset.
\newblock {\em Computer optics}, 40(2):294--300, 2016.

\bibitem{simonyan2014very}
Karen Simonyan and Andrew Zisserman.
\newblock Very deep convolutional networks for large-scale image recognition.
\newblock {\em arXiv preprint arXiv:1409.1556}, 2014.

\bibitem{stallkamp2011german}
Johannes Stallkamp, Marc Schlipsing, Jan Salmen, and Christian Igel.
\newblock The german traffic sign recognition benchmark: a multi-class classification competition.
\newblock In {\em The 2011 international joint conference on neural networks}, pages 1453--1460. IEEE, 2011.

\bibitem{szegedy2015going}
Christian Szegedy, Wei Liu, Yangqing Jia, Pierre Sermanet, Scott Reed, Dragomir Anguelov, Dumitru Erhan, Vincent Vanhoucke, and Andrew Rabinovich.
\newblock Going deeper with convolutions.
\newblock In {\em Proceedings of the IEEE conference on computer vision and pattern recognition}, pages 1--9, 2015.

\bibitem{szegedy2016rethinking}
Christian Szegedy, Vincent Vanhoucke, Sergey Ioffe, Jon Shlens, and Zbigniew Wojna.
\newblock Rethinking the inception architecture for computer vision.
\newblock In {\em Proceedings of the IEEE conference on computer vision and pattern recognition}, pages 2818--2826, 2016.

\bibitem{tu2020empirical}
Lifu Tu, Garima Lalwani, Spandana Gella, and He~He.
\newblock An empirical study on robustness to spurious correlations using pre-trained language models.
\newblock {\em Transactions of the Association for Computational Linguistics}, 8:621--633, 2020.

\bibitem{wang2021scaled}
Chien-Yao Wang, Alexey Bochkovskiy, and Hong-Yuan~Mark Liao.
\newblock Scaled-yolov4: Scaling cross stage partial network.
\newblock In {\em Proceedings of the IEEE/cvf conference on computer vision and pattern recognition}, pages 13029--13038, 2021.

\bibitem{xie2016traffic}
Kaixuan Xie, Shiming Ge, Qiting Ye, and Zhao Luo.
\newblock Traffic sign recognition based on attribute-refinement cascaded convolutional neural networks.
\newblock In {\em Advances in Multimedia Information Processing-PCM 2016: 17th Pacific-Rim Conference on Multimedia, Xi{\'{}} an, China, September 15-16, 2016, Proceedings, Part I}, pages 201--210. Springer, 2016.

\bibitem{yang2022improving}
Zhuolin Yang, Zhikuan Zhao, Boxin Wang, Jiawei Zhang, Linyi Li, Hengzhi Pei, Bojan Karla{\v{s}}, Ji~Liu, Heng Guo, Ce~Zhang, et~al.
\newblock Improving certified robustness via statistical learning with logical reasoning.
\newblock {\em Advances in Neural Information Processing Systems}, 35:34859--34873, 2022.

\bibitem{yu2020bdd100k}
Fisher Yu, Haofeng Chen, Xin Wang, Wenqi Xian, Yingying Chen, Fangchen Liu, Vashisht Madhavan, and Trevor Darrell.
\newblock Bdd100k: A diverse driving dataset for heterogeneous multitask learning.
\newblock In {\em Proceedings of the IEEE/CVF conference on computer vision and pattern recognition}, pages 2636--2645, 2020.

\bibitem{zhang2022cctsdb}
Jianming Zhang, Xin Zou, Li-Dan Kuang, Jin Wang, R~Simon Sherratt, and Xioafeng Yu.
\newblock Cctsdb 2021: a more comprehensive traffic sign detection benchmark.
\newblock {\em Human-centric Computing and Information Sciences}, 12, 2022.

\bibitem{zhang2022care}
Jiawei Zhang, Linyi Li, Ce~Zhang, and Bo~Li.
\newblock Care: Certifiably robust learning with reasoning via variational inference.
\newblock {\em arXiv preprint arXiv:2209.05055}, 2022.

\bibitem{zheng2021graph}
Qinkai Zheng, Xu~Zou, Yuxiao Dong, Yukuo Cen, Da~Yin, Jiarong Xu, Yang Yang, and Jie Tang.
\newblock Graph robustness benchmark: Benchmarking the adversarial robustness of graph machine learning.
\newblock {\em arXiv preprint arXiv:2111.04314}, 2021.

\bibitem{zhu2016traffic}
Zhe Zhu, Dun Liang, Songhai Zhang, Xiaolei Huang, Baoli Li, and Shimin Hu.
\newblock Traffic-sign detection and classification in the wild.
\newblock In {\em Proceedings of the IEEE conference on computer vision and pattern recognition}, pages 2110--2118, 2016.

\end{thebibliography}
\end{document}